\documentclass[a4paper,twocolumn,showpacs,
nofootinbib,superscriptaddress]{revtex4-2}

\usepackage{newpxtext,newpxmath}
\usepackage{bm}
\usepackage{tensor}
\usepackage{float}
\usepackage{dcolumn}
\usepackage{cancel}
\usepackage[colorlinks]{hyperref}
\usepackage[usenames,dvipsnames]{color}
\hypersetup{
     breaklinks=true,
    pdfstartview={FitH},    
    colorlinks=true,       
    linkcolor=blue,          
    citecolor=red,        
    filecolor=magenta,      
    urlcolor=blue,           
    anchorcolor=green,      
    linktocpage=true
}

\usepackage{verbatim}

\usepackage{graphicx}
\usepackage{epsfig}
\usepackage{xcolor}

\newcommand{\pder}[2]{\dfrac{\partial#1}{\partial#2}}
\newcommand{\de}{\mathrm{d}}
\newcommand{\lin}{\\[7pt]}
\newcommand{\red}{\color{red!60!black}}

\newcommand{\beqa}{\begin{eqnarray}}
\newcommand{\eeqa}{\end{eqnarray}}
\newcommand{\p}{\partial}
\newcommand{\nn}{\nonumber}
\newcommand{\bs}{\boldsymbol}

\def\div{{\rm div}}

\begin{document}

\title{Dark Gravitational Sectors  on a 
Generalized Scalar-Tensor Vector Bundle Model and Cosmological Applications}

\author{Spyros Konitopoulos}
\email{spykoni@gmail.com}
\affiliation{N.C.S.R. Demokritos, Institute of Nuclear and Particle Physics, 
Patr. Gregoriou E \& 27 Neapoleos Str, 
15341 Agia Paraskevi, Athens, Greece}

\author{Emmanuel N. Saridakis}
\email{msaridak@phys.uoa.gr}
\affiliation{Department of Physics, National Technical University of Athens, 
Zografou
	Campus GR 157 73, Athens, Greece}
\affiliation{National Observatory of Athens, Lofos Nymfon, 11852 Athens, 
	Greece}
\affiliation{Department of Astronomy, School of Physical Sciences, University 
	of Science 
	and Technology of China, Hefei 230026, P.R. China}

\author{P. C. Stavrinos}
\email{pstavrin@math.uoa.gr}
\affiliation{Department of Mathematics, National and Kapodistrian University of 
Athens,
	Panepistimiopolis 15784, Athens, Greece}

\author{A. Triantafyllopoulos}
\email{alktrian@phys.uoa.gr}
\affiliation{Section of Astrophysics, Astronomy and Mechanics, Department of 
Physics,  
	National and Kapodistrian University of Athens, Panepistimiopolis 15784,
	Athens, Greece}

\begin{abstract}
In this work we present the foundations of generalized scalar-tensor theories 
arising from vector bundle constructions,  and we study the kinematic, 
dynamical and cosmological consequences. In particular, over a 
pseudo-Riemannian space-time base manifold, we define a fibre structure with 
two scalar fields. The resulting space is a 6-dimensional vector bundle  
endowed with a non-linear connection. We provide the form of the geodesics and 
the   Raychaudhuri and general field equations, both in Palatini and metrical 
method. When applied at a cosmological framework, this novel 
geometrical structure  induces extra terms in the modified Friedmann equations, 
leading to the appearance of an effective dark energy sector, as well as of an 
interaction of the dark mater sector with the metric. We show that   we can 
obtain the standard thermal history of the universe, with the sequence of matter 
and dark-energy epochs, and furthermore  the effective dark-energy 
equation-of-state parameter can lie in the quintessence or phantom regimes, or 
exhibit the phantom-divide crossing.

\end{abstract}


\maketitle

\section{Introduction}

Modified gravity has attracted a large amount of research for two 
reasons and thus motivations. Firstly, at the purely theoretical level, it 
improves the renormalizability of General Relativity and 
hence it may be the first step towards gravitational quantization 
\cite{Stelle:1976gc}. Secondly, at the phenomenological, cosmological, level, 
it is one of the two main ways that can 
offer an  explanation for the the early- and late-time accelerated phases of 
the expansion of the universe
\cite{Nojiri:2006ri,Clifton:2011jh}. Hence, it has an advantage 
comparing to the alternative way, which is to introduce by hand the  
  inflaton or/and dark energy sectors while maintaining General Relativity as 
the underlying gravitational theory 
\cite{Copeland:2006wr,Cai:2009zp}.

Modified gravity theories can be obtained as extensions of the
Einstein-Hilbert  Lagrangian through the addition of extra terms, such as in   
 $f(R)$ gravity  \cite{Starobinsky:1980te,Capozziello:2002rd}, in $f(G)$
gravity \cite{Nojiri:2005jg}, in  Weyl gravity \cite{Mannheim:1988dj}, in 
  Lovelock gravity \cite{Lovelock:1971yv}, etc. Additionally, they can be 
obtained through the insertion of extra scalar fields, coupled with curvature 
invariants, such as in the general  class of scalar-tensor theories  
\cite{Horndeski:1974wa,Nicolis:2008in,Deffayet:2009mn,Bamba:2015uxa}.
However, one interesting class of modified gravity arises from the 
consideration of alternative geometries, beyond the Riemannian framework of 
General Relativity. Thus, one can  start from the equivalent, torsional 
formulation of gravity and extend it obtaining  
$f(T)$ gravity \cite{Cai:2015emx}, $f(T,T_G)$ gravity
\cite{Kofinas:2014owa}, etc. Similarly, one can allow for non-metricity, 
obtaining symmetric teleparallel gravity  \cite{BeltranJimenez:2017tkd}, $f(Q)$ 
gravity \cite{Jimenez:2019ovq}, etc.

Inspired by the above, one may proceed to the construction of gravitational 
modifications through a more radical modification of the underlying 
geometrical structure, 
namely considering Finsler or Finsler-like geometries
\cite{Miron,Vacaru:2005ht,kour-stath-st2012,Triantafyllopoulos:2018bli,Minas:2019urp,
Kouretsis:2010vs,Mavromatos:2010jt,
Basilakos:2013hua,Basilakos:2013ij,
Chang:2007vq,Vacaru:2010fi,Vacaru:2010rd,Kostelecky:2008be,
Foster:2015yta,Kostelecky:2011qz,Pfeifer:2011xi,Kostelecky:2012ac,Stavrinos:2012ty,Stavrinos:2021ygh,
Hohmann:2016pyt,Hohmann:2018rpp,Pfeifer:2019wus,Hohmann:2020mgs}.
In the framework of these generalized metric structures 
in a vector bundle, scalar-tensor theories can naturally appear, and in 
particular  the scalar fields  play the role of fibres or internal 
variables \cite{stavrinos-ikeda1999, stavrinos-ikeda2000, stavrinos-alexiou, 
Ikeda:2019ckp}. 

On the other hand, theoretical and observational cosmological evidence have 
indicated the existence of dark matter sector
\cite{Farnes:2017gbf,Hsueh:2019ynk,Capozziello:2006uv,Peebles:2000yy,Moreschi:2015qya,DiValentino:2019jae,Borowiec:2006qr,Capozziello:2008it,Alvarenga:2016yxh,Arbey:2006it,Relancio:2020zok}. Based on observational results, 
dark matter plays a significant role in the    evolution  of the universe, 
especially concerning the growth of structures \cite{Davis:1985rj}. 
Additionally, since its microphysics is unknown one could have the interesting 
case in which dark matter interacts with dark energy \cite{Farrar:2003uw}, a case 
that has significant advantages since it can lead to the alleviation of the 
coincidence 
problem  \cite{Zimdahl:2001ar} as well as of the $H_0$ tension 
\cite{DiValentino:2020zio}. Hence, 
 the investigation of dark sectors in   modified theories of 
gravity and cosmology is a 
fundamental subject for cosmological phenomena.

In the present work we are interested in constructing
  Finsler-like geometrical structures, which will induce scalar-tensor theory 
with two scalars-fibres models. In particular, we   consider
 a pseudo-Riemannian 4-dimensional
space-time with two fibres and we   investigate 
the properties of 
$F^6$ space-time, with non-holonomic structures,
extracting the 
Raychaudhuri and field equations. Finally, we apply these geometrical 
generalized scalar-tensor theories on vector bundle constructions on a 
cosmological framework, in order to examine their cosmological implications on 
the effective dark energy and dark matter sectors.

The paper is organised as follows.  In Section \ref{section 2}  we present the
basic geometrical concepts of the theory, analyzing the metric 
decomposition and the appearance of the geometric dark sectors, investigating 
also 
the geodesic structure. In Section \ref{Field Equations}  we consider the 
action on the fibre bundle, we derive 
the field equations with both Palatini and metrical methods, in holonomic and 
non-holonomic forms, and finally construct the involved energy-momentum 
tensor, incorporating the contributions of the dark matter sector.
In Section \ref{Raychaudhuri}
we examine the 
Raychaudhuri equations in the context of the $F^6$ bundle geometry.
In Section \ref{Cosmology} we proceed to the application on a 
cosmological framework, showing the appearance of an effective dark sector that has a purely geometrical origin and which can lead to a universe behavior in agreement 
with observations.
Finally, in Section \ref{Conclusions} we discuss the concluding remarks.

\section{Scalar-tensor theories induced from the Vector Bundle}
\label{section 2}

In this section we present the basics of the geometrical framework under 
consideration
\cite{stavrinos-ikeda1999, stavrinos-ikeda2000, stavrinos-alexiou, 
Ikeda:2019ckp}. Firstly we will 
review the basic structure of the Lorentz   fibre bundle, then we will 
describe the metric splitting and the appearance of the geometric dark sectors, 
and finally we will proceed to the geodesic investigation.

\subsection{Basic structure of the Lorentz scalar tensor fibre bundle}

We consider a 4-dimensional manifold $M$ equipped with coordinates 
$x^\mu$, $\mu=0,\ldots,3$ and a Lorentzian metric $g_{\mu\nu}(x)$ with 
signature 
$(-,+,+,+)$ on it. Over any open subset of $M$ we define a fibre structure with 
two scalar degrees of freedom $\phi^{(1)}$ and $\phi^{(2)}$. The resulting 
space 
is a 6 dimensional space-time fibre bundle, $F^6$, over the pseudo-Riemannian 
base manifold $M$, with local coordinates $\{U^M\} = \{x^\mu, \phi^a\}$, which 
trivializes locally to the product, $M \times \{\phi^{(1)}\} \times 
\{\phi^{(2)}\}$. Capital indices $K,L,M,N,Z\ldots$ span all the range of values 
of indices on a fibre bundle's tangent space.
Additionally, a coordinate transformation on the fibre bundle maps the old 
coordinates to the new as:
\begin{align}
	x^{\mu} \mapsto & x^{\prime\mu}(x^\nu) \label{x-transormation} \\
	\phi^{a}(x) \mapsto & \phi^{\prime a}(x^\prime) = \delta^{a}_{b}\phi^{b}(x)
\end{align}
where $ \delta^{a}_{b} $ is the Kronecker symbol for the corresponding latin 
indices $a, b$ which take values in the range $\{(1), (2)\}$
and the Jacobian matrix $\pder{x^{\prime\mu}}{x^\nu}$ is non-degenerate.

In the  space at hand, the adapted basis is defined as
\begin{equation}\label{adapted basis}
	\left\{X_M\right\} = \left\{\delta_\mu,  \p_{(1)}, \p_{(2)} \right\}
\end{equation}
where
\beqa\label{horizontal basis}
\delta_\mu = \partial_\mu - N^{(1)}_\mu(x^\nu, 
\phi^a)\partial_{(1)}-N^{(2)}_\mu(x^\nu, \phi^a)\partial_{(2)}
\eeqa
with 
$\partial_\mu \equiv
\pder{}{x^\mu}$
and $ \partial_{a} \equiv \pder{}{\phi^{a}} $.
The fields $N^{a}_\mu(x^\nu, \phi^b)$ comprise a special type of nonlinear
connection and it is a fundamental structure of the framework under 
consideration, since it connects the base manifold's tangent space with the one 
of the fibre. Furthermore, the dual basis is
$
	\{X^M\} =\{\de x^\mu,
	\delta\phi^{(1)}, \delta\phi^{(2)}
	\}
$
where
$
\delta\phi^{a} = \de\phi^{a} + N^{a}_\mu(x^\nu, \phi^b)\de 
	x^\mu
$
and
$a=1,2$.  Finally, the basis vectors transform as:
\begin{equation}
	\delta^\prime_{\mu} = \pder{x^\nu}{x^{\prime\mu}}\delta_\nu~~, ~~
	\partial^{\,\prime}_a = \delta^{b}_{a} \partial_{b}
\end{equation}
where summations are implied over the ranges of values of $\mu$ and $a$.

From its defining relations \eqref{adapted basis},
\eqref{horizontal basis}, the non commutative nature of the adapted basis can 
be easily revealed. Specifically we obtain  
\beqa\label{adapted base algebra}
\mathcal [X_M,X_N]=\tensor{\mathcal W}{^L_{MN}}X_L
\eeqa
where $\tensor{\mathcal W}{^L_{MN}}$ are the structure functions 
of the adapted base algebra
which obey the Jacobi identity\footnote{$ \circlearrowleft_{M,N,L} $
indicates summation with respect to the cyclic permutation of the indices 
$M,N,L$.},
\beqa
\circlearrowleft_{M,N,L}
\{
X_M\tensor{\mathcal W}{^R_{NL}}+
\tensor{\mathcal W}{^R_{MS}}\tensor{\mathcal W}{^S_{NL}}
\}=0
\eeqa
As can be directly observed, the non-zero components of the structure functions 
are,
\beqa\label{structure functions components}
\tensor{\mathcal W}{^L_{MN}}=\{
\tensor{\tilde W}{^a_{\mu\nu}},
\tensor{W}{^a_{\mu b}}
\}
\eeqa
where 
\beqa
\tensor{\tilde W}{^a_{\mu\nu}}&=&\delta_\nu N^a_\mu-\delta_\mu N^a_\nu\nn \\
\tensor{W}{^a_{\mu b}}&=&\p_b N^a_\mu
\eeqa

The metric structure of the fibre bundle is defined as
\begin{equation}\label{metric}
	\pmb{G} = g_{\mu\nu}(x)\,\de x^\mu\otimes\de x^\nu + 
	v_{ab}(x)\,\delta\phi^{a}\otimes \delta\phi^{b}
\end{equation}
Furthermore, the form of the fibre metric is assumed to be
\begin{equation}\label{vdef}
	v_{ab}(x) = \delta_{ab}\phi(x)
\end{equation}
and transforms as $
	v^\prime_{ab}(x^\prime) = \delta^{c}_{a}\delta^d_b v_{cd}(x)
$.
This particular choice \eqref{vdef} encodes the mutual independence of the fibre scalar fields and their equivalent contribution in the internal space geometry.

The covariant derivative of a base vector $X_M$ over E, 
with respect to a base vector $X_N$, is in general
\beqa
D_{X_N}X_M=\tensor{\pmb{\Gamma}}{^L_{MN}}X_L
\eeqa
A special connection structure is chosen \cite{Ikeda:2019ckp}, such that the 
non-vanishing 
components of the vector bundle connection are\footnote{Note that the selected 
connection structure is not the usual d-connection which preserves by 
parallelism the horizontal and vertical distributions \cite{Miron}.
} 
\beqa\label{connection structure}
\tensor{\pmb{\Gamma}}{^L_{MN}}=\{
\tensor{L}{^\lambda_{\mu\nu}}, \tensor{L}{^c_{a\nu}}, \tensor{C}{^c_{\mu b}}, 
\tensor{C}{^{\lambda}_{ab}}\}
\eeqa
The above local connections determine the action of the covariant derivatives 
upon the adapted basis of the bundle.
Further details about the geometrical structure of our consideration are given in 
Appendix \ref{Connection and curvature}.

Alongside with the general symmetry property
$
\tensor{\bs\Gamma} {^L_{[MN]}}=0
$
and under a trivial permutation of the indices, the general metricity condition
\beqa\label{metricity condition}
D_{X_M}\pmb{G}=0
\eeqa
leads to the result
\begin{equation}\label{Symmetric Connection}
    \tensor{\bs\Gamma}{^L_{MN}} = \frac{1}{2}\mathcal G^{RL} \left( X_M \mathcal G_{NR} + X_N 
\mathcal G_{RM} - X_R \mathcal G_{MN}\right)
\end{equation}
As will soon be illustrated, relation \eqref{Symmetric Connection}
does not 
imply a Levi-Civita tensor. The non-holonomic nature of the adapted basis 
\eqref{adapted base algebra} gives rise to torsion contribitions (see relation 
\eqref{torsion bundle}).
Taking into account the presumed special connection structure 
\eqref{connection structure} we arrive at the following explicit expressions 
for 
the non-vanishing components of the vector bundle special, linear connection:
\beqa
\label{Levi-civita connection}
&&\tensor{L}{^\mu_{\nu\lambda}}(x)=\tensor{\Gamma}{^\mu_{\nu\lambda}}(x) \\
\label{special connection 1}
&&\tensor{C}{^a_{\mu b}}=\tensor{L}{^a_{b\mu}}=
\delta^a_b\frac{1}{2\phi}\partial_\mu\phi \\
\label{special connection 2}
&&\tensor{C}{^\mu_{ab}}=-\frac{1}{2}\delta_{ab}g^{\mu\nu}\partial_\nu\phi
\eeqa
where $\tensor{\Gamma}{^\mu}_{\nu\lambda}$ is the Levi-Civita connection 
of the second kind\footnote{It is obvious from \eqref{structure functions 
components} that the structure
functions $\tensor{\mathcal W}{^L_{MN}}$ nullify 
if all indices are space-time. Therefore, they do not add torsion if restricted 
in the base manifold.}.  

The curvature tensor of a linear connection is defined as
\begin{align}\label{Riemann}
\tensor{\mathcal R}{^K_{LMN}}= & X_M \tensor{\bs\Gamma}{^K_{LN}} - X_N 
\tensor{\bs\Gamma}{^K_{LM}} +
\tensor{\bs\Gamma}{^R_{LN}} \tensor{\bs\Gamma}{^K_{RM}}- 
	 \tensor{\bs\Gamma}{^R_{LM}} \tensor{\bs\Gamma}{^K_{RN}}
	 \nonumber \\ & - \tensor{\mathcal W}{^R_{MN}} \tensor{\bs\Gamma}{^K_{LR}}
\end{align}
In the holonomic base limit, $\tensor{\mathcal W}{^L_{MN}}=0$, the generalised 
curvature tensor \eqref{Riemann} reduces to the standard Riemann tensor.

The torsion tensor of the vector bundle is defined as
\beqa\label{torsion general}
\tensor{\mathscr T}{^L_{MN}}=2\tensor{\bs\Gamma}{^L_{[MN]}}+
\tensor{\mathcal W}{^L_{MN}}
\eeqa
Since in our case $\tensor{\bs\Gamma}{^L_{[MN]}}=0$, we have
\beqa\label{torsion bundle}
\tensor{\mathscr T}{^L_{MN}}=
\tensor{\mathcal W}{^L_{MN}}
\eeqa
Analogously, we define the generalized Ricci tensor
\beqa
\tensor{\mathcal R}
{_{MN}}
&\equiv&
\tensor{\mathcal G}
{^L_K}
\tensor{\mathcal R}
{^K_{MLN}}=
\tensor{\mathcal R}
{^L_{MLN}}
\nn \\
&=&\tensor{X}{_L}\tensor{\bs\Gamma}{^L_{MN}}-
\tensor{X}{_N}\tensor{\bs\Gamma}{^L_{ML}}+\tensor{\bs\Gamma}{^L_{MN}}\tensor{
\bs\Gamma}{^R_{LR}}-
\tensor{\bs\Gamma}{^L_{MR}}\tensor{\bs\Gamma}{^R_{LN}}\nn \\
&&+\tensor{\bs\Gamma}{^L_{MR}}
\tensor{\mathcal W}{^R_{NL}}
\eeqa
The last term casts the tensor non-symmetric as can be directly seen in 
\eqref{Ricci antisymmetric}.

For the linear connection \eqref{Levi-civita connection}-\eqref{special 
connection 2} we obtain the non-zero components of the generalised Ricci 
tensor\footnote{Note that as is obvious from \eqref{Ricci antisymmetric} the 
generalised Ricci tensor is non-symmetric. Despite the fact that $\mathcal 
R_{\mu a}=0$, we see that $\mathcal R_{a\mu}\ne 0$}
\beqa\label{Ricci-greek}
\mathcal R_{\mu\nu} &=&
 R_{\mu\nu} + \tensor{R}{^{(\phi)}_{\mu\nu}} \\
\label{Ricci-latin}
\mathcal R_{ab}&=& -\frac{1}{2}\square\phi 
\delta_{ab}+\frac{1}{2}\delta_{ac}(\p^\lambda\phi)
\tensor{W}{^c_{\lambda b}} \\
\label{Ricci-cross}
\mathcal R_{a\mu}&=&
\tensor{C}{^\nu_{ab}}
\tensor{\tilde W}{^b_{\mu\nu}}
\eeqa
where $\square \equiv D^\mu D_\mu$, $R_{\mu\nu}$ is the Ricci tensor of 
Levi-Civita connection and
\beqa
\tensor{R}{^{(\phi)}_{\mu\nu}}=
\frac{1}{2\phi^2}
\partial_\mu\phi\partial_\nu\phi-
\frac{1}{\phi}
D_\mu D_\nu\phi
+\frac{1}{2\phi}\partial_\mu\phi
\tensor{W}{^a_{\nu a}}
\eeqa
the contribution of the pure scalar field.

Multiplying \eqref{special connection 2} with $v^{ab}$ we can
express the quantity $\p^\mu\phi$ in terms of $C^\mu_{~ab}$. Indeed, it is easy
to see that
\beqa\label{partial-connection}
\p^\mu\phi=-\phi v^{ab}\tensor{C}{^\mu_{ab}}=-\delta^{ab}\tensor{C}{^\mu_{ab}}
\eeqa

The corresponding scalar curvature is
\beqa\label{fibre scalar curvature}
\mathcal R&=&g^{\mu\nu}\mathcal R_{\mu\nu} + 
v^{ab}\mathcal R_{ab}=R+
R^{(\phi)}
\eeqa
where $R$ is the Levi-Civita curvature and with the aid of \eqref{partial-connection},
\beqa
R^{(\phi)}=-\frac{2}{\phi}\square\phi-v^{ab}\left(
\frac{1}{2\phi}\p_\mu\phi+
\tensor{W}{^c_{\mu c}}
\right)\tensor{C}{^\mu_{ab}}
\eeqa

Lastly, the generalised Einstein's tensor is
\beqa
\label{Einsten Tensor generalised}
\tensor{\mathcal E}{_{MN}}=\mathcal R_{MN}-\frac{1}{2}\mathcal R \mathcal 
G_{MN}
\eeqa
The tensor $\tensor{\mathcal E}{_{MN}}$ contains extra 
terms that come from the introduction of internal variables $\phi^{(1)}, 
\phi^{(2)}$ and their derivatives, giving a possible locally anisotropic 
contribution. 

\subsection {The Geometrical Effects of Dark Gravitational Field}

In the previous subsection we presented the underlying geometrical structure of 
the scalar-tensor theories that are induced from the Vector Bundle. Hence, we 
can now proceed to the investigation of their effects on the physical  
quantities such as the metric, and in particular of the appearance of dark 
sectors.

In order to  account for the effects of the geometry of 
space-time on dark sectors, we follow the general study elaborated in 
\cite{Savvopoulos:2020}. In particular, the metric $g_{\mu\nu}(x)$ of the 
base manifold $M$ is assumed to decompose into an ``ordinary'' $(O)$ and a 
``dark'' matter sector $(D)$
\beqa\label{base metric separation}
g_{\mu\nu}(x) = \tensor{g}{^{(O)}_{\mu\nu}}(x) + \tensor{g}{^{(D)}_{\mu\nu}}(x)
\eeqa
since from a physical point of view a unified description of gravity may 
include 
the gravitational interaction of both \cite{PandaXII:2018}.
In the following, we postulate that the fibre space remains unaffected by the 
dark sector.

In analogy with \eqref{base metric separation},
the Levi-Civita connection admits contributions from the ordinary $(O)$ and 
dark matter $(D)$ energy densities alongside a term that arises from their 
mutual interaction:
\beqa\label{Christoffel 2nd kind}
\tensor{\Gamma}{^\mu_{\nu\lambda}}(x)&=&\tensor{\Gamma}{^{(O)\mu}_{\nu\lambda}}
(x)+
\tensor{\Gamma}{^{(D)\mu}_{\nu\lambda}}(x)+\tensor{\gamma}{^{\mu}_{\nu\lambda}}
(x)
\eeqa
Substituting \eqref{Christoffel 2nd kind} in the definition relation $
\tensor{\Gamma}{^\mu_{\nu\lambda}}=g^{\mu\rho}\Gamma_{\rho\nu\lambda}
$,
the interaction part 
$\tensor{\gamma}{^\mu_{\nu\lambda}}$ can be easily expressed in 
terms of the 
inverse of the total, ordinary and dark matter metrics, as well as the 
respective connection parts of the second kind, namely
\begin{equation}\label{gamma interaction}
\tensor{\gamma}{^\mu_{\nu\lambda}}=
\left(
g^{\mu\rho}-g^{(O)\mu\rho}
\right)\tensor{\Gamma}{^{(O)}_{\rho\nu\lambda}}+
\left(
g^{\mu\rho}-g^{(D)\mu\rho}
\right)\tensor{\Gamma}{^{(D)}_{\rho\nu\lambda}}
\end{equation}
As it is evident from \eqref{partial-connection} the connection 
$\tensor{C}{^\mu_{ab}}$ depends linearly on the inverse of the space-time 
metric. 
Therefore, it should split in a manner similar to \eqref{Christoffel 2nd kind}, 
i.e.
\beqa
\tensor{C}{^\mu_{ab}}(x)=\tensor{C}{^{(O)\mu}_{ab}}(x)+\tensor{C}{^{(D)\mu}_{ab}
}(x)+\tensor{c}{^{\mu}_{ab}}(x)
\eeqa
where 
\beqa
\tensor{C}{^{(O)\mu}_{ab}}&=&-\frac{1}{2}\delta_{ab}g^{(O)\mu\nu}
\partial_\nu\phi\nn \\
\tensor{C}{^{(D)\mu}_{ab}}&=&-\frac{1}{2}\delta_{ab}g^{(D)\mu\nu}
\partial_\nu\phi
\eeqa
and 
\beqa
\tensor{c}{^\mu_{ab}}=-\frac{1}{2}\delta_{ab}\left(
g^{\mu\nu}-g^{(O)\mu\nu}-g^{(D)\mu\nu}\right)\partial_\nu\phi
\eeqa
All other connections are not conditioned in such splittings, since, as can be 
seen from \eqref{special connection 1},
they are not directly related to the metric of the base manifold.

In order to make manifest the contributions of the ordinary, dark matter, 
scalar 
and interaction sectors in the Ricci tensor, let us reformulate the above 
expressions in a spirit analogous to \cite{Savvopoulos:2020}. We 
have, 
\beqa\label{Riemanian Ricci}
R_{\mu\nu} &=& \tensor{R}{^{(O)}_{\mu\nu}}+
\tensor{R}{^{(D)}_{\mu\nu}}+
r_{\mu\nu}
\eeqa
where
$r_{\mu\nu}$ expresses the
interactions between ordinary and dark matter.

In similar lines, we assume an analogous splitting for the $\square$ operator, 
namely
\beqa
\square\phi=\left(\square^{(O)}+\square^{(D)}+{\scriptstyle\square}\right)\phi
\eeqa
where
\beqa
\square\phi&=&g^{\mu\nu}D_\mu D_\nu\phi=g^{\mu\nu}\left(\p_\mu\p_\nu\phi-
\tensor{\Gamma}{^\lambda_{\mu\nu}}\p_\lambda\phi\right)\nn \\
\square^{(O)}\phi&=&g^{(O)\mu\nu}\left(\p_\mu\p_\nu\phi-\tensor{\Gamma}
{^{(O)\lambda}_{\mu\nu}}\p_\lambda\phi\right)\nn \\
\square^{(D)}\phi&=&g^{(D)\mu\nu}\left(\p_\mu\p_\nu\phi-\tensor{\Gamma}
{^{(D)\lambda}_{\mu\nu}}\p_\lambda\phi\right)
\eeqa
Accordingly to \eqref{gamma interaction}, the interaction part ${\scriptstyle\square}$ can be expressed in 
terms of
Christofell symbols and of the inverse of the total, ordinary and dark matter metrics.

Considering all the above, we can now write 
\beqa
\mathcal R_{ab}=\tensor{\mathcal R}{^{(O)}_{ab}}+
\tensor{\mathcal 
R}{^{(D)}_{ab}}+\mathfrak{r}_{ab}
\eeqa
where
\beqa
\tensor{\mathcal R}{^{(O)}_{ab}}&=&-\frac{1}{2}\delta_{ab}\square^{(O)}\phi-
\frac{1}{2}
\delta_{ac}\delta^{de}
\tensor{W}{^c_{\mu b}}
\tensor{C}{^{(O)\mu}_{de}} \\ \tensor{\mathcal 
R}{^{(D)}_{ab}}&=&-\frac{1}{2}\delta_{ab}\square^{(D)}\phi-
\frac{1}{2}
\delta_{ac}\delta^{de}\tensor{W}{^c_{\mu b}}
\tensor{C}{^{(D)\mu}_{de}}\nn \\
\mathfrak{r}_{ab}&=&
-\frac{1}{2}\delta_{ab}{\scriptstyle\square}\phi-
\frac{1}{2}
\delta_{ac}\delta^{de}\tensor{W}{^c_{\mu b}}
\tensor{c}{^\mu_{de}}
\eeqa
Analogously with the above, the extra fibre 
contribution to the Ricci scalar and the Einstein tensor can be 
straightforwardly decomposed into ordinary, dark and interaction sectors.

\subsection{Geodesics}

We close this section with an investigation of the geodesic structure of the 
theory. In particular, we will derive the geodesic equations imposing the 
auto-parallel condition
on the vector tangent to the geodesic curve.
Let
\beqa
Y=Y^\mu \delta_\mu + Y^a \p_a
\eeqa
be the tangent vector. Then from the auto-parallel condition
$
D_Y Y=0
$
we obtain the pair of geodesic equations
\beqa\label{1st geodesic}
&&\frac{d^2x^\mu}{d\tau^2}+
\tensor{\Gamma}{^\mu_{\nu\lambda}}
\frac{dx^\nu}{d\tau}
\frac{dx^\lambda}{d\tau}
+
\tensor{C}{^\mu_{ab}}
\frac{\delta\phi^a}{d\tau}
\frac{\delta\phi^b}{d\tau}=0
\eeqa
\beqa\label{2nd geodesic}
\frac{d}{d\tau}
\left(
\frac{\delta\phi^a}{d\tau}
\right)+
\tensor{L}{^a_{b\mu}}
\frac{dx^\mu}{d\tau}
\frac{\delta\phi^b}{d\tau}
=0
\eeqa
where
\beqa
\label{geodesic parameter}
\frac{d}{d\tau}
\equiv 
\frac{dx^\mu}{d\tau}
\delta_\mu+
\frac{\delta\phi^a}{d\tau}
\p_a=Y
\eeqa
Multiplying \eqref{1st geodesic} with the mass of a test particle and 
inserting \eqref{Christoffel 2nd kind}, we can reveal the kinematic influence 
of 
each of the sectors of our geometrical structure. Indeed we acquire
\beqa
&&
\!\!\!\!\!\!\!\!\!
m\left(
\frac{d^2x^\mu}{d\tau^2}
+\tensor{\Gamma}{^{(O)\mu}_{\nu\lambda}}
\frac{dx^\nu}{d\tau}
\frac{dx^\lambda}{d\tau}
\right)\nn \\
&&
\!\!\!\!\!
=
-m\left(
\tensor{\Gamma}{^{(D)\mu}_{\nu\lambda}}+\tensor{\gamma}{^\mu_{\nu\lambda}}
\right)
\frac{dx^\nu}{d\tau}
\frac{dx^\lambda}{d\tau}
-m\tensor{C}{^\mu_{ab}}
\frac{\delta\phi^a}{d\tau}
\frac{\delta\phi^b}{d\tau}\nn\\
\eeqa
The three terms that appear on the right hand side of the above equation
account for the deviation from Riemannian geometry. This deviation reflects the 
presence of dark matter and its interaction with the ordinary sector and reveal 
the influence of the hidden scalar fields on the motion 
of particles. From the 
point of view of an observer who does not take into account the existence of 
these hidden entities, the three terms on the right are interpreted as inertial 
forces.

Substituting \eqref{special connection 1} and \eqref{special connection 2}, 
into \eqref{1st geodesic} and \eqref{2nd geodesic} we obtain
\beqa
\label{geodesic1 sub}
&&
\frac{d^2x^\mu}{d\tau^2}
+\tensor{\Gamma}{^\mu_{\nu\lambda}}
\frac{dx^\nu}{d\tau}
\frac{dx^\lambda}{d\tau}
-\frac{1}{2}
\delta_{ab}\p^\mu\phi
\frac{\delta\phi^a}{d\tau}
\frac{\delta\phi^b}{d\tau}
=0\nn \\ \\
\label{geodesic2 sub}
&&
\frac{d}{d\tau}
\left(
\frac{\delta\phi^a}{d\tau}
\right)+
\frac{1}{\phi}
\p_\mu\phi
\frac{dx^\mu}{d\tau}
\frac{\delta\phi^a}{d\tau}
=0
\eeqa
It can be easily verified that \eqref{geodesic2 sub} has the exact 
solution 
\beqa
\frac{\delta\phi^a}{d\tau}
=\frac{C^a}{\phi}
\eeqa
where $C^a$ are constants of integration. 
Inserting the above solution into \eqref{geodesic1 sub} leads to
\beqa
\frac{d^2x^\mu}{d\tau^2}
+\tensor{\Gamma}{^\mu_{\nu\lambda}}
\frac{dx^\nu}{d\tau}
\frac{dx^\lambda}{d\tau}
-\frac{1}{2}
\p^\mu\phi
\frac{C^2}{\phi^2}
=0
\eeqa
where $C^2=(C^{(1)})^2+(C^{(2)})^2$ and $\tensor{\Gamma}{^\mu_{\nu\lambda}}$ is 
given in \eqref{Christoffel 2nd kind}.

Additionally, it is   instructive to examine separately the special case 
where the geodesics for the Riemannian part is given by  
\beqa\label{geodesics 1}
\frac{d^2x^\mu}{d\tau^2} + \tensor{\Gamma}{^\mu_{\kappa\lambda}} 
	\frac{dx^\kappa}{d\tau}\frac{dx^\lambda}{d\tau} = 0
\eeqa
and for the internal structure by 
\beqa
\label{parallelism condition}
\frac{\delta\phi^a}{d\tau}=0
\eeqa
or equivalently,
\beqa\label{geodesics 2}
\frac{d\phi^{a}}{d\tau} = - N^{a}_\mu \frac{dx^\mu}{d\tau}
\eeqa
In our model the form of the geodesics is given by both the relations 
\eqref{geodesics 1}, \eqref{geodesics 2}. The non-linear connection $N^{a}_\mu$ 
interconnects
the differential of the internal quantity $\phi^a$ with the velocity of the 
observer. Such an interconnection can be interpreted as 
a manifestation of a 
condition of parallelism \eqref{parallelism condition}. 

Lastly, note that considering a specific form for the non-linear connection, 
for instance
\begin{equation}\label{NLconnection}
N_\mu^{a} = \frac{A(\phi)}{2\phi} \partial_\mu\phi\, \phi^{a}
\end{equation}
equation \eqref{geodesics 2} has the solution
\begin{equation}
\phi^a(x)=\phi_0^a(x)e^{-\frac{1}{2}\int_0^\tau A(\phi)d(\ln\phi)}
\end{equation}
where $\phi_0^a(x)=\phi^a(x)|_{\tau=0}$.

\section{Field Equations}
\label{Field Equations}

In the previous section we presented the geometric structure and the kinematic 
variables of the examined construction. In the present section we proceed to
physics. In particular, we consider the action on the fibre bundle, we derive 
the field equations with both Palatini and metrical methods, we examine the 
Raychaudhuri equations, and we finally construct the involved energy-momentum 
tensor.

The total action of the theory is
\begin{align}\label{Fields Action}
	S & = S_G + 2\kappa S_M \nonumber\\
	  & = \int_Q d^6 U \sqrt{|\mathcal G|} \mathcal{G}^{AB} 
	  \mathcal R_{AB} + 2\kappa \int_Q d^6 U \sqrt{|\mathcal G|} \mathcal L_M
\end{align}
where $\mathcal L_M (\mathcal G^{MN}, \Psi^i)$ is the matter   Lagrangian, 
$\Psi^i$
the various matter fields described collectively, and 
$Q$ a closed subspace of $F^6$.
For additional details we refer to the Appendix \ref{sec: Field Eqs}.

\subsection{Palatini method}

Firstly, we follow the Palatini method in which the variation is performed 
independently for the fields $\mathcal G_{AB}$ and $\pmb{\Gamma}^L_{~MN}$ (see 
Appendix \ref{sec: Field Eqs}). 
If we assume a metrical compatible connection we acquire
\begin{equation}\label{Field Eqs Palatini 1}
	\mathcal R_{(MN)} - \frac{1}{2}\mathcal G_{MN} \mathcal R = \kappa \mathcal 
T_{MN}
\end{equation}
and
\begin{equation}\label{Field Eqs Palatini 2}
	\mathcal G^{MN} \tensor{\mathscr T}{^A_{KA}} + \mathcal G^{ML}
	\left( \tensor{\mathscr T}{^N_{LK}} - \tensor{\mathscr 
T}{^A_{LA}}\delta^N_K \right) = 0
\end{equation}
where $\tensor{\mathscr T}{^K_{MN}}$ is the torsion of the connection, 
given in \eqref{torsion general}, and specifically in our case 
\eqref{torsion bundle}. The coupling constant $\kappa$ in \eqref{Field Eqs Palatini 1} will be determined in the General Relativity (GR) limit of the theory.
As is evident 
from 
\eqref{structure functions components}, the only independent nonzero components 
of the torsion tensor are the following:
\begin{align}
	\tensor{\mathscr T}{^a_{\lambda b}} & = \tensor{W}{^a_{\lambda b}} = 
\partial_b N^a_\lambda \nonumber\\
	\tensor{\mathscr T}{^a_{\lambda\nu}} & = 
	\tensor{\tilde W}{^a_{\lambda\nu}} = \delta_\nu N^a_\lambda - 
\delta_\lambda N^a_\nu
\end{align}
From these expressions, as well as \eqref{Field Eqs Palatini 2} we acquire
\begin{align}
   \tensor{W}{^a_{ \lambda b}} & = 0 \\
   \tensor{\tilde W}{^a_{\lambda\nu}} & = 0
\end{align}
i.e. we find that all the torsion components vanish. We see that the Palatini
field equations force the connection to coincide with the Levi-Civita 
connection, in total agreement with the Levi-Civita theorem.
Therefore, if one wishes to study a non-holonomic structure of the adapted 
basis, one has to abandon the Palatini method of variation. 
Nevertheless, let us 
continue the study and analyse the field equations \eqref{Field Eqs Palatini 
1}. 
On the spacetime manifold we have
\begin{align}
\label{Field Eqs Palatini 1 Sasaki}
	& E_{\mu\nu} + \frac{1}{\phi}g_{\mu\nu} \left[\square\phi - 
	\frac{1}{4\phi} 
	\partial^\lambda\phi\partial_\lambda\phi\right] - \frac{1}{\phi}D_\mu 
D_\nu\phi 
	\nonumber\\
	& \ \ \ \ \  + \frac{1}{2\phi^2} \partial_\mu\phi \partial_\nu 
	\phi = 8\pi G \mathcal T_{\mu\nu}
\end{align}
while on the fibre 
\beqa \label{Field Eqs Palatini 2 Sasaki}
\left(
-R+\frac{1}{\phi}\square\phi-\frac{1}{2\phi^2}\p_\mu\phi\p^\mu\phi
\right)v_{ab}=16\pi G \mathcal T_{ab}
\eeqa
where $E_{\mu\nu}$ is the standard Einstein's tensor of GR while the extra terms in \eqref{Field Eqs Palatini 1 Sasaki} come from the spacetime components of the generalized tensor \eqref{Einsten Tensor generalised}. One can recover the standard Einstein field equations of GR from \eqref{Field Eqs Palatini 1 Sasaki} in the limit $\partial_\mu \phi \rightarrow 0$, in which the coupling constant is determined as $\kappa = 8\pi G$, with $G$ the Newtonian gravitational constant. We mention that the energy-momentum tensor corresponding to the Lagrangian of 
the matter fields
$\mathcal L_M(\mathcal G^{MN},\Psi^i)$ is defined in the standard way. 
Specifically, we have
\begin{equation}
	\mathcal T_{\mu\nu} = -\dfrac{2}{\sqrt{-g}}\dfrac{\delta(\sqrt{-g}\mathcal 
L_M)}{\delta g^{\mu\nu}}
\end{equation}
for its space-time components, and
\begin{equation}
	\mathcal T_{ab} = -\dfrac{2}{\sqrt{v}}\dfrac{\delta(\sqrt{v}\mathcal 
L_M)}{\delta v^{ab}}
\end{equation}
for its fibre components.

In summary, from relations \eqref{Field Eqs Palatini 1 Sasaki}, 
\eqref{Field Eqs Palatini 2 Sasaki}
we deduce that the field equations include additional terms because of fibre 
fields and dark matter considerations.

\subsection{Metrical method}

In this subsection we proceed to the extraction of the field equations 
following the metrical method. In particular,  
  we will derive the bundle field equations by varying the action 
  \eqref{Fields 
Action}  with only respect to the metric $\mathcal G^{MN}$. Doing so, we obtain 
(see Appendix \ref{sec: Field Eqs}):
\beqa\label{Metric Field Eqs Bundle}
&&
\tensor{\mathcal E}{_{(MN)}}\nn \\
+&&\left(
\tensor{\delta}{^L_{(M}}
\tensor{\delta}{^R_{N)}}-
\tensor{\mathcal G}{^{LR}}
\tensor{\mathcal G}{_{MN}}
\right)
\left(
\tensor{D}{_{L}}
\tensor{\mathcal W}{^A_{RA}}-
\tensor{\mathcal W}{^B_{LB}}\tensor{\mathcal W}{^C_{RC}}
\right)\nn \\&& = \kappa\tensor{\mathcal T}{_{MN}}
\eeqa
The fields of curvature and torsion must obey the Bianchi identities 
\eqref{Bianchi General 1},
\eqref{Bianchi General 2}. Specifically, the first identity takes the form (see 
Appendix \ref{sec: Bianchi}):
\beqa	 
 D^A\mathcal E_{AN}+\tensor{\mathcal R}{^A_R}
\tensor{\mathcal W}{^R_{NA}}+
\frac{1}{2}\tensor{\mathcal R}{^{KA}_{NR}}
\tensor{\mathcal W}{^R_{AK}}=0
\eeqa
In order to derive a generalization of the continuity equation we isolate the
symmetric part of the tensor $\mathcal E_{AN}$. Employing \eqref{Ricci 
antisymmetric} we write,
\beqa\label{Bianchi diff}
&&D^A\mathcal E_{(AN)}+\tensor{\mathcal R}{^A_R}
\tensor{\mathcal W}{^R_{NA}}+
\frac{1}{2}\tensor{\mathcal R}{^{KA}_{NR}}
\tensor{\mathcal W}{^R_{AK}}\nn \\
&&+\frac{1}{2}D^A\left(
\tensor{\mathcal W}{^L_{RA}}
\tensor{\pmb{\Gamma}}{^R_{LN}}-
\tensor{\mathcal W}{^L_{RN}}
\tensor{\pmb{\Gamma}}{^R_{LA}}
\right)
=0
\eeqa
Now, from \eqref{Metric Field Eqs Bundle} we see that,
\beqa\label{Field Equation diff}
D^A\mathcal E_{(AN)}+
D^AH_{AN}=Q_N
\eeqa
where 
\beqa
H_{AN}&=&\tensor{\Delta}{^{LR}_{AN}}\left(
D_L\tensor{\mathcal W}{^K_{RK}}-
\tensor{\mathcal W}{^K_{LK}}
\tensor{\mathcal W}{^S_{RS}}
\right)\nn \\
\tensor{\Delta}{^{LR}_{AN}}&=&
\frac{1}{2}\left(
\tensor{\mathcal G}{^L_{A}}
\tensor{\mathcal G}{^R_{N}}+
\tensor{\mathcal G}{^L_{N}}
\tensor{\mathcal G}{^R_{A}}
\right)-
\tensor{\mathcal G}{^{LR}}
\tensor{\mathcal G}{_{AN}}\nn \\
Q_N&=&\kappa D^A\tensor{\mathcal T}{_{AN}} \label{H Delta and Q definitions}
\eeqa
Thus, inserting \eqref{Bianchi diff} into \eqref{Field Equation diff}
we arrive at a final expression for the dissipation vector, namely
\beqa
Q_N&=&\tensor{\mathcal R}{^A_R}
\tensor{\mathcal W}{^R_{AN}}+
\frac{1}{2}\tensor{\mathcal R}{^{AK}_{NR}}
\tensor{\mathcal W}{^R_{AK}}\nn \\
&&+\frac{1}{2}D^A\Big[
\tensor{\mathcal W}{^L_{RN}}
\tensor{\pmb{\Gamma}}{^R_{LA}}-
\tensor{\mathcal W}{^L_{RA}}
\tensor{\pmb{\Gamma}}{^R_{LN}}+\nn \\
&&+2\tensor{\Delta}{^{LR}_{AN}}
\left(
D_L\tensor{\mathcal W}{^K_{RK}}-
\tensor{\mathcal W}{^K_{LK}}
\tensor{\mathcal W}{^S_{RS}}
\right)
\Big] \label{dissipation vector}
\eeqa
As it is evident from the above expression, the conservation of energy is 
restored, namely $Q_N=0$,
when the torsions $\tensor{\mathcal W}{^L_{MN}}$ are set to zero.

In terms of the space-time and fibre components the generalised field equation 
\eqref{Metric Field Eqs Bundle}
respectively reads,
\begin{align}
	&\mathcal E_{(\mu\nu)}  + \left(\delta^{\lambda}_{(\mu} \delta^{\rho}_{\nu)} 
- g^{\lambda\rho}g_{\mu\nu}\right) \left( D_\lambda \tensor{W}{^c_{\rho c}} - 
\tensor{W}{^d_{\lambda d}} \tensor{W}{^c_{\rho c}}\right)\nn \\
	& \quad +
	v^{ab}g_{\mu\nu}\tensor{C}{^\lambda_{ab}}\tensor{W}{^c_{\lambda c}}
	= \kappa \mathcal T_{\mu\nu} \label{Field Eqs Metric 1}\\
	&\mathcal E_{(ab)}
	- g^{\lambda\rho}v_{ab} \left( D_\lambda \tensor{W}{^c_{\rho c}} - 
\tensor{W}{^d_{\lambda d}} \tensor{W}{^c_{\rho c}} \right) \nn \\
	& \quad + \tensor{C}{^\lambda_{ab}}\tensor{W}{^c_{\lambda c}}
= \kappa \mathcal T_{ab} \label{Field Eqs Metric 2}
\end{align}
where $\mathcal E_{\mu\nu}$ and $\mathcal E_{ab}$ are the spacetime and fibre components respectively of the generalized Einstein's tensor \eqref{Einsten Tensor generalised}. These equations must reproduce general relativity in the appropriate limit. We 
find that for $\dot\phi\rightarrow 0$ and $\mathcal W^K_{MN} \rightarrow 0$ 
Eqs. 
\eqref{Field Eqs Metric 1} reduce to the Einstein field equations of GR for the 
metric 
$g_{\mu\nu}$, provided that the coupling constant takes the value $\kappa = 8\pi G$, where $G$ is the Newtonian gravitational constant. In general, the value of $\kappa$ depends on the structure of the geometry. Moreover, in this limit, Eq. \eqref{Field Eqs Metric 2} gives the condition:
\begin{equation}
    \overset{H}{\mathcal T} = \overset{V}{\mathcal T} \label{EM gr limit}
\end{equation}
with
\[\overset{H}{\mathcal T}=g^{\mu\nu}\mathcal T_{\mu\nu}=
\mathcal T^\mu_{~~~\mu}
\]
and
\[\overset{V}{\mathcal T}=
v^{ab}\mathcal T_{ab}=\mathcal T^a_{~~~a}
\]
i.e. the traces of the spacetime energy momentum $ \overset{H}{\mathcal T} $ 
and 
of the fibre one $ 
\overset{V}{\mathcal T} $ are equal in the GR limit. 

From \eqref{energy momentum tensor} and assuming that the matter
Lagrangian $\mathcal L_M$ depends on the metric $\mathcal G_{MN}$ but not on 
its derivatives we acquire:
\begin{align}
    \mathcal T_{\mu\nu} & = -2\pder{\mathcal L_M}{g^{\mu\nu}} + \mathcal L_M 
g_{\mu\nu} \label{Tmn(L)}\\
    \mathcal T_{ab} & = -2\pder{\mathcal L_M}{v^{ab}} + \mathcal L_M v_{ab} 
\label{Tab(L)}
\end{align}
From \eqref{EM gr limit}, \eqref{Tmn(L)} and \eqref{Tab(L)} we obtain the GR 
limit condition for the matter Lagrangian:
\begin{equation}\label{matter lagrangian gr limit condition}
    v^{ab}\pder{\mathcal L_M}{v^{ab}} = -\mathcal L_M + 
g^{\mu\nu}\pder{\mathcal 
L_M}{g^{\mu\nu}}
\end{equation}
In this limit and for a matter fluid with a barotropic equation of state 
$P_m(\rho^{(0)}) $ and a conserved current 
$D_\mu\left(\rho^{(0)}Y^\mu\right)=0$, with $\rho^{(0)}$ the rest mass energy 
density, the energy-momentum tensor reads \cite{Minazzoli:2012md}:
\begin{equation}\label{EM(L)}
    \mathcal T^{\mu\nu} = -\rho^{(0)}\frac{\partial\mathcal 
L_M}{\partial\rho^{(0)}}Y^\mu Y^\nu + \left(\mathcal L_M
\frac{\partial\mathcal 
L_M}{\partial\rho^{(0)}} - \rho^{(0)} \right)g^{\mu\nu}
\end{equation}
where the following relation has been used:
\begin{equation}
    \frac{d\rho^{(0)}}{d g^{\mu\nu}} = \frac{1}{2} \rho^{(0)} \left( g_{\mu\nu} 
+ Y_\mu Y_\nu \right)
\end{equation}
Comparison of \eqref{EM(L)} with \eqref{perf fluid} gives
\begin{equation}
    \frac{\partial\mathcal L_M}{\partial\rho^{(0)}} = -\frac{\rho_m + 
    P_m}{\rho^{(0)}}\quad, \quad \mathcal L_M = -\rho_m \label{matter 
lagrangian}
\end{equation}
Finally from \eqref{matter lagrangian} and \eqref{matter lagrangian gr limit 
condition} we obtain
\begin{align}
     v^{ab}\pder{\mathcal L_M}{v^{ab}} & = \rho_m + g^{\mu\nu}
     \frac{\partial\mathcal L_M}{\partial\rho^{(0)}} \frac{d\rho^{(0)}}{d 
g^{\mu\nu}}\Leftrightarrow \nonumber\\
     \frac{2}{\phi}\pder{\rho_m}{\phi} & = \frac{\rho_m}{2} + \frac{3P_m}{2}
\end{align}
This equation determines the dependence of the barotropic fluid's energy density $\rho_m$ on the scalar field $\phi$ at the GR limit.

\subsection {Incorporation of Dark Matter in Energy-Momentum Tensor}

We close this section by discussing the energy-momentum tensor. 
The theory at hand allows for
two sources of dark matter. A purely geometrical one,
in which dark matter is attributed to the effective properties of the bundle 
structure, 
and   a fluid/particle one in which   dark matter contributes
directly to the energy momentum tensor. 

Following the geometrical method, we re-arrange the terms of
\eqref{Field Eqs 
Metric 1}, so 
that only 
the standard GR Einstein's tensor appears on the lhs:
\beqa
E_{\mu\nu}=\kappa\tilde{\mathcal T}_{\mu\nu}
\eeqa
where 
\beqa\label{total energy-momentum tensor}
\tilde{\mathcal T}_{\mu\nu}&=&
\mathcal T_{\mu\nu}+\tensor{\mathcal T}{^{(\phi)}_{\mu\nu}}
\eeqa
and
\beqa
&&\tensor{\mathcal T}{^{(\phi)}_{\mu\nu}}=\nn \\
&&-\frac{1}{\kappa}
\Big[
\tensor{E}{^{\phi}_{(\mu\nu)}}+
\left(\delta^{\kappa}_{(\mu} \delta^{\lambda}_{\nu)} - 
g^{\kappa\lambda}g_{\mu\nu}\right) \left( D_\kappa \tensor{W}{^a_{\lambda a}} - 
\tensor{W}{^b_{\kappa b}} \tensor{W}{^c_{\lambda c}}\right)\nn \\
&&~~~~~~~~
+v^{ab}g_{\mu\nu}\tensor{C}{^\lambda_{ab}}\tensor{W}{^c_{\lambda c}}
\Big]
\eeqa
Hence,  the geometrical properties of our model can be viewed as additional 
terms to the energy momentum tensor and therefore, in the GR framework, 
interpreted as effective dark matter.

In addition to this, one can directly include dark matter contributions to the 
energy momentum tensor
 \cite{Savvopoulos:2020},
\beqa
\mathcal T_{\mu\nu}=
\tensor{\mathcal T}{^{(O)}_{\mu\nu}}+
\tensor{\mathcal T}{^{(D)}_{\mu\nu}}+
\tensor{\tau}{_{\mu\nu}} \label{Tmn split}
\eeqa
so that \eqref{total energy-momentum tensor} becomes
\beqa\label{energy momemtum spacetime total}
\tilde{\mathcal T}_{\mu\nu}=
\tensor{\mathcal T}{^{(O)}_{\mu\nu}}+
\tensor{\mathcal T}{^{(D)}_{\mu\nu}}
+\tensor{\mathcal T}{^{(\phi)}_{\mu\nu}}+\tensor{\tau}{_{\mu\nu}}
\eeqa
The above sectorial decomposition of the energy-momentum tensor induces the 
corresponding decomposition of the generalised Einstein's tensor.
From the above relation we notice that the total form of the energy momentum tensor $\tilde{\mathcal T}_{\mu\nu}$ includes the fibre contributions as well as the dark matter sector and its interactions with ordinary matter. It is possible that a conformal relation between ordinary and dark matter exists \cite{Capozziello:2008it}.

\section{Raychaudhuri Equations}
\label{Raychaudhuri}
It is known that the
Raychaudhuri's equations describe the evolution of the acceleration of the universe through the gravitating fluid. Their form depends on the metrical structure of space, i.e. in spaces with generalized metric structure and torsion as in a Finsler space-time \cite{stavrinos-alexiou}.
The Raychaudhuri's equations are produced by the deviation of nearby geodesics or fluid lines and monitor their evolution. In our case, they are twofold extended. On the one hand, with the introduction of the scalars $\phi^{(1)}$, $\phi^{(2)}$ and on the other, with the inclusion of the dark sector.

In order to examine the local behaviour of a single, 
time-like geodesic among the congruence, let us assume the tangent vector,
\beqa
Y^M\equiv\left(
\frac{dx^\mu}{d\tau}, \frac{\delta\phi^a}{d\tau}
\right)
\eeqa
which
satisfies the auto-parallel condition along the track of the geodesic 
\beqa\label{autaoparallel Y}
D_Y Y=0
\eeqa
Furthermore, we assume that $\tau$ is properly chosen in order for $Y^M$
to have a unit norm\footnote{
This assumption is consistent with the definition of the geodesic parameter, 
\eqref{geodesic parameter}. It does not alter the signature of the
Riemannian metric, and forces the extra fibre variables to behave as space-like 
components. The fact that the extra degrees of freedom do not transform 
covariantly is not incompatible with the existence of a co-moving observer in 
the bundle $F^6$.
}, namely
\beqa\label{tangent norm}
\mathcal G_{MN}Y^M Y^N=-1
\eeqa
The 2nd rank tensor  
\beqa
\tensor{\mathcal B}{^M_N}=D_N Y^M=
X_N Y^M+
\tensor{\Gamma}
{^M_{LN}}Y^L
\eeqa
measures the failure of the separation vector between adjacent geodesics to be 
parallelly transported 
along the congruence \cite{Wald, Carroll}. 
From the auto-parallel condition, we obtain
\beqa
Y^N \mathcal B_{MN}=0
\eeqa
and from \eqref{tangent norm}  we get
\beqa
D_M\left(Y_N Y^N\right)=0\Rightarrow Y^M \mathcal B_{MN}=0
\eeqa
Additionally, we can separate the space part of the metric, making use of the projective tensor $\mathcal H_{MN}$,
\beqa
\mathcal G_{MN}=\mathcal H_{MN}-Y_M Y_N
\eeqa
Indeed, it is easy to see that, 
\beqa
\mathcal H_{MN}Y^N=0
\eeqa
As a 2nd rank tensor, $\mathcal B_{MN}$ can be decomposed into its irreducible 
components, namely its trace, traceless symmetric and antisymmetric part. 
In particular, the trace of the tensor $\mathcal B_{MN}$ is called 
\emph{expansion},
i.e.
\begin{equation}\label{Expansion}
\Theta=\mathcal G^{ML}\mathcal G^{NR}\mathcal B_{MN}\mathcal H_{LR}=\mathcal 
B^{MN}\mathcal H_{MN}=
\tensor{\mathcal B}
{^M_M}= D_M Y^M
\end{equation}
and is a measure
of the volume change of a sphere of test particles centered on the geodesic.
The symmetric, traceless part of the same tensor is called \emph{shear}:
\beqa\label{Shear}
\mathcal S_{MN}=\mathcal B_{(MN)}-\frac{1}{5}\Theta \mathcal H_{MN}
\eeqa
and describes the shape distortion of the test particles from the initial sphere 
to an ellipsoid. Lastly, the antisymmetric part of the tensor is called
\emph{rotation}
\beqa\label{Rotation}
\Omega_{MN}=\mathcal B_{[MN]}
\eeqa
and describes the rotation of the initial sphere of test particles. 

Now, the 2nd rank tensor $\mathcal B_{MN}$ can be written in terms of its 
irreducible components as  
\beqa
\mathcal B_{MN}=\frac{1}{5}\Theta \mathcal H_{MN}+
\mathcal S_{MN}+\Omega_{MN}
\eeqa
Taking into account
\eqref{autaoparallel Y},
the definition of the Riemann tensor \eqref{Riemann} and the fact that 
\beqa
[D_L, D_N]Y_M=
\tensor{\mathcal W}{^R_{LN}}  D_R Y_M - \tensor{\mathcal R}{^{R}_{MLN}}Y_R
\eeqa
we obtain that the covariant derivative of $\mathcal B_{MN}$
along the geodesic is 
\begin{equation}
Y^L  D_L\mathcal B_{MN}= \tensor{\mathcal W}{^R_{LN}}Y^L \mathcal B_{MR}-
\tensor{\mathcal R}{^R_{MLN}}Y^L Y_R-
\tensor{\mathcal B}{^L_N}\mathcal B_{ML}
\end{equation}
Taking the trace of the above equation we result to
\begin{equation}
\frac{d\Theta}{d\tau}=
\tensor{\mathcal W}{^L_{MN}}Y^M
\tensor{\mathcal B}{^N_L}-\mathcal R_{MN}Y^M Y^N-\mathcal B^{MN}\mathcal B_{NM}
\end{equation}
Written in terms of the irreducible components of $\mathcal B_{MN}$, the above 
equation provides the extension of the {\it Raychaudhuri equation} on a general 
space-time vector bundle, namely
\beqa\label{Raychadhuri generalisation}
\frac{d\Theta}{d\tau}&=&
\tensor{
\mathcal W}{^L_{MN}}Y^M\tensor{
\mathcal B}{^N_L}-\mathcal R_{MN}Y^M Y^N\nn \\&&-\frac{1}{5}\Theta^2-\mathcal 
S^{MN}\mathcal S_{MN}+\Omega^{MN}\Omega_{MN}
\eeqa
For the specific choice of special connection structure \eqref{connection 
structure} we acquire
\beqa
\label{generalised expansion}
\Theta&=&\div Y+\frac{d}{d\tau}\left[
\ln\left(\sqrt{-\mathcal G}\right)
\right]=\theta+\theta^{(\phi)}
\eeqa
where $\div Y=X_M Y^M$, $\mathcal G=\phi^2 g$ is
the determinant of the bundle metric, and
\beqa
\label{expansion space-time}
\theta&=&
\nabla_\mu Y^\mu=
\p_\mu Y^\mu+\frac{d}{d\tau}\left[\ln{\sqrt{-g}}\right] \\
\label{expansion fibre}
\theta^{(\phi)}&=&
\p_a Y^a-
N^a_\mu
\p_a Y^\mu+\frac{d}{d\tau}\left(\ln{\phi}
\right)
\eeqa
To the standard expansion $\theta$, a contribution of purely geometric origin $\theta^{(\phi)}$ is added. It is produced by the the scalars $\phi^{(a)}$, the non-linear connection $N^a_\mu$ and the fibre components of the tangent vector $Y^a$. The form and the overall sign of this contribution 
\eqref{expansion fibre} is directly related to the kinematics of the universal 
evolution and under certain circumstances it provides a triggering inflation 
mechanism. Especially, in the case of an inflaton scalar field, the contribution of the volume $\theta^{(\phi)}$ will be positive and an increase of volume can appear. 

In the same manner we can calculate each of the terms of \eqref{Raychadhuri 
generalisation}. For the non-holonomic term we obtain
\beqa
\tensor{
\mathcal W}{^L_{MN}}Y^M\tensor{
\mathcal B}{^N_L}&=&
\tensor{\tilde W}{^a_{\mu\nu}}Y^\mu D_a Y^\nu\nn \\&&+
\tensor{W}{^a_{\mu b}}\left(
Y^\mu D_a Y^b-
Y^b D_a Y^\mu\right) \ \ 
\eeqa
The generalised tidal term decomposes into its Riemannian part plus the 
additional contributions that rise from the additional geometric structure  
\beqa\label{tidal}
\mathcal R_{MN}Y^M Y^N
&=&R_{\mu\nu}Y^\mu Y^\nu+
\tensor{R}{^{(\phi)}_{\mu\nu}}Y^\mu Y^\nu+\mathcal R_{ab}Y^a Y^b\nn \\
&&+
\tensor{C}{^\nu_{ab}}
\tensor{\tilde W}{^b_{\mu\nu}}
Y^a Y^\mu
\eeqa
where
\beqa
	\mathcal S_{\mu\nu}&=&
	\sigma_{\mu\nu}+\tensor{S}{^{(\phi)}_{\mu\nu}}\nn \\
	\mathcal S_{ab}&=&\frac{1}{2}\left(
	\p_b Y_a+\p_a Y_b
	-2\tensor{C}{^\mu_{ab}}Y_\mu
	\right)-\frac{1}{5}\Theta \mathcal H_{ab}\nn \\
	\mathcal S_{\mu a}&=&
	\left(
	\p_a Y_\mu+\delta_\mu Y_a-2\tensor{C}{^b_{\mu a}}Y_b
	\right)-
	\frac{1}{5}\Theta \mathcal H_{\mu a}
	\eeqa
and
	\beqa
	\sigma_{\mu\nu}&=&\nabla_{(\nu} Y_{\mu )}-\frac{1}{3}\theta\mathcal 
H_{\mu\nu}\nn \\
	\tensor{S}{^{(\phi)}_{\mu\nu}}&=&\frac{1}{15}\left(
	2\theta-3\theta^{(\phi)}
	\right)\mathcal H_{\mu\nu}-\frac{1}{2}\left(
	N^a_\mu\p_a Y_\nu+
	N^a_\nu\p_a Y_\mu
	\right)\nn \\
	\eeqa
Finally, we can acquire a similar decompositions for the generalised rotation 
too, namely
	\beqa
	\Omega_{\mu\nu}&=&
	\omega_{\mu\nu}+
	\tensor{\Omega}{^{(\phi)}_{\mu\nu}}
	\nn \\
	\Omega_{ab}&=&
	\frac{1}{2}\left(
	\p_b Y_a-\p_a Y_b
	\right)\nn \\
	\Omega_{\mu a}&=&
	\frac{1}{2}\left(
	\p_a Y_\mu-\delta_\mu Y_a
	\right)
	\eeqa
	where
	\beqa
	\omega_{\mu\nu}&=&
	\nabla_{[\nu}Y_{\mu]}\nn \\
	\tensor{\Omega}{^{(\phi)}_{\mu\nu}}&=&
	\frac{1}{2}\left(
	N^a_\mu\p_a Y_\nu-
	N^a_\nu\p_a Y_\mu
	\right)
	\eeqa
Assembling all the pieces together we obtain 
\begin{equation}
\label{generalised Raychaudhuri}
\frac{d}{d\tau}\left(\theta + \theta^{(\phi)}\right)
=-R_{\mu\nu}Y^\mu Y^\nu
	-\frac{1}{3}\theta^2-\sigma^{\mu\nu}\sigma_{\mu\nu}+
	\omega^{\mu\nu}\omega_{\mu\nu} + \mathscr Q
	\end{equation}
	with
	\beqa\label{additional contribution in Raychaudhuri}
	\mathscr Q & = & \tensor{\tilde W}{^a_{\mu\nu}}Y^\mu\p_a Y^\nu+
	\tensor{W}{^a_{\mu b}}\left(
	Y^\mu D_a Y_b-Y_b D_a Y^\mu
	\right)\nn \\
	&&-\tensor{R}{^{(\phi)}_{\mu\nu}}Y^\mu Y^\nu-\mathcal R_{ab}Y^a Y^b\nn \\
	&&+\frac{2}{15}\theta^2-\frac{1}{5}\left[
	2\theta\theta^{(\phi)}+\left(\theta^{(\phi)}\right)^2
	\right]\nn \\
&&-S^{(\phi)\mu\nu}\tensor{S}{^{(\phi)}_{\mu\nu}}-2\sigma^{\mu\nu}\tensor{S}{^{
(\phi)}_{\mu\nu}}\nn \\
	&&-S^{ab}S_{ab}-2S^{\mu a}S_{\mu a}\nn \\
	&&+\Omega^{(\phi)\mu\nu}
	\tensor{\Omega}{^{(\phi)}_{\mu\nu}}
	+2\omega^{\mu\nu}
	\tensor{\Omega}{^{(\phi)}_{\mu\nu}}\nn \\
	&&+\Omega^{ab}\Omega_{ab}+2\Omega^{\mu a}\Omega_{\mu a}
	\eeqa
As we can see from \eqref{generalised Raychaudhuri} $\mathscr Q$
disturbs the rate of the volume change for a number of reasons. Firstly because of the interaction between the volumes $\theta$
and $\theta^{(\phi)}$, secondly due to the contribution of the scalar fields $\phi^{(a)}$ and 
lastly because of the presence of of the non-linear connection $N^a_\mu$ and the torsion functions $\tensor{\tilde{W}}{^a_{\mu\nu}}$, $\tensor{W}{^a_{\mu b}}$.

The generalised tidal field \eqref{tidal} includes the standard Riemann 
contribution \eqref{Riemanian Ricci} and additional terms which can affect the 
evolution of the gravitational fluid for possible singularities/conjugate points 
in the universe. It is obvious, because of extra internal geometrical concepts 
of fibre-fields $\phi(x)$, of the non-linear connection $N^a_\mu$ in the 
metrical structure of our model $F^6$ and of the introduction of the dark 
gravitational field.
In the framework of our space, $F^6$ and for a given congruence of time-like 
geodesics, the expansion $\Theta$, shear $S_{\mu\nu}$ and rotation $\Omega$ are 
described, in a generalised form, in Eqs
\eqref{Expansion}, \eqref{Shear}, \eqref{Rotation} which provide us the 
generalised type of Raychaudhuri equation \eqref{generalised Raychaudhuri}. The 
extra terms affect the variation of the volume during the evolution of fluid 
lines (focusing/defocusing) in the accelerating expansion of the universe. This 
is possible due to the perturbation of the deviation equation of nearby 
geodesics or curves. 

In a comoving frame, the term in Eq. \eqref{additional contribution in 
Raychaudhuri} involving the structure functions $\tensor{\tilde W}{^a_{\mu\nu}}$ 
vanishes. As we will see later, this is in agreement with the generalized 
Friedmann equations for this model. Specifically, in those equations, in which 
the matter fluid is at rest, no such term appears. This is an important test for 
the consistency of this model because the generalized Raychaudhuri equation 
\eqref{generalised Raychaudhuri} should not give additional information on the 
kinematics of the FRW comoving frame.
It is worthwhile to mention that for a constant non-linear connection the above 
equations can be drastically simplified. 

With the aid of the bundle field equations \eqref{Metric Field Eqs Bundle} we 
acquire
\beqa\label{Einstein-Raychaudhuri}
&&
\tensor{\mathcal R}{_{MN}}Y^M Y^N\nn \\
&&+\left(
\tensor{D}{_{M}}
\tensor{\mathcal W}{^A_{NA}}-
\tensor{\mathcal W}{^B_{MB}}\tensor{\mathcal W}{^C_{NC}}
\right)
\left(
Y^M Y^N-\frac{1}{4}
\tensor{\mathcal G}{^{MN}}
\right)
\nn \\
&&=\kappa\left(
\tensor{\mathcal T}{_{MN}}Y^MY^N+\frac{1}{4}\mathcal T
\right)
\eeqa
where $\mathcal T=\tensor{\mathcal T}{_{MN}}\tensor{\mathcal G}{^{MN}}$.
The sign of the generalised tidal field  determines the evolution of the volume 
of the fluid lines. It is evident that it does not only depend on the energy and 
current density of matter, but also on the structure of the algebra of the 
adapted basis.

\section{Cosmology with non-linear Connection}
\label{Cosmology}In the previous sections we presented the geometric formalism 
in which 
generalised scalar-tensor theories and dark gravitational sectors are induced 
from the vector bundle. In this section we proceed to the explicit cosmological 
application of such constructions.

In order to construct a cosmological framework, we need to extend the standard spatially homogeneous and isotropic Friedmann-Robertson-Walker (FRW)  metric of ordinary Riemannian geometry and GR on the fibre bundle $E$. In particular, we consider the flat case of the former $(k=0)$, as the simplest one in GR, and extend it to account for the additional degrees of freedom of $E$ in the following way:
\begin{align}
	\mathbf{G} = & \, -\de t\otimes\de t  + a^2(t)\left(\de x\otimes\de x + \de 
	y\otimes\de y + \de z\otimes\de z\right) \nonumber \\
	& \, + \phi(t)\left(\delta\phi^{(1)}\otimes\delta\phi^{(1)} + 
	\delta\phi^{(2)}\otimes\delta\phi^{(2)}\right) \label{bundle metric}
\end{align}
We mention here that  the observational 
constraints on the  (almost zero) spatial curvature have been extracted under the 
consideration of the usual FRW metric in Riemannian geometry, and thus in 
principle one cannot deduce that the same feature would necessarily hold in the 
case of the present extended geometric structure. Nevertheless, since in our 
work we are interested in performing a first cosmological application, we 
impose zero spatial curvature.
As one can see, the first line of \eqref{bundle metric} is the standard 
4-dimensional
spatially flat   FRW   metric,
while 
the 
second line arises from the additional structure of the Lorentz fibre bundle. The additional degrees of freedom of the metric as well as the anholonomicity of the adapted basis are expected to enrich the dynamics of space-time, compared to the standard spatially isotropic and flat FRW cosmology \cite{Vacaru:2002}\footnote{To examine whether the symmetries of the standard FRW solution persist, a careful and meaningful definition of these symmetries should be given in the current framework of extended space-time. The most consistent way to do this is by means of Lie derivatives and extended Killing vectors on the bundle $E$ or by direct implementation of the method of complete lifts \cite{Pfeifer:2011xi}. Using these tools, we could construct spatially homogeneous and isotropic cosmological solutions that may even extrapolate the classification into spatially flat, closed or open. This would be an interesting topic for a future project.}

Moreover, we consider the matter sector to correspond to a perfect fluid, with 
 energy-momentum tensor  of the form
\begin{equation} \label{perf fluid}
	\mathcal T_{\mu\nu} = (\rho_m+P_m)Y_\mu Y_\nu + P_m g_{\mu\nu}
\end{equation}
with $\rho_m$ is the energy density, $P_m$ the pressure and $Y^\mu$ the bulk 
4-velocity of the fluid.

We will first study the equations derived from the metrical method, since the 
Palatini equations occur as a special case of the former. For the spacetime 
\eqref{bundle metric}, and with the perfect fluid \eqref{perf fluid},  the 
non-diagonal components of the field equations \eqref{Field Eqs Metric 1}, 
\eqref{Field Eqs Metric 2} give
\begin{eqnarray}
   &&  \tensor{\tilde W}{^a_{0i}}\dot\phi = 0 ,\nonumber\\
   &&
     \left[
     \tensor {W}{^{(1)}_{0(2)}} + \tensor {W}{^{(2)}_{0(1)}}
     \right]
     \dot\phi = 0 \nonumber\\
     &&
     \tensor{ W}{^a_{ia}} \tensor {W}{^b_{jb}} = 0
     \label{Einstein Eqs nondiagonal 1}
\end{eqnarray}
for $i\neq j$, and
\begin{equation}\label{Einstein Eqs nondiagonal 2}
   \tensor {W}{^a_{ia}} \left(\frac{\dot\phi}{4\phi} - H - \tensor {W}{^b_{0b}} 
\right) = 0
\end{equation}
where   $0$ stands for the coordinate time component, $i,j = 1,2,3$ for the 
spatial
components, $a,b = (1),(2)$ for the fibre components, and a dot denotes 
differentiation 
with respect to  time: $\dot\phi = \frac{d\phi}{dt}$.
Furthermore from \eqref{Einstein Eqs nondiagonal 1} and the spatial isotropy 
of \eqref{bundle metric} and assuming that $\dot\phi \neq 0$, we acquire $$\tensor 
{W}{^a_{ia}} = 0 = \tilde W^a_{0i}
$$ and
$$\tensor{W}{^{(1)}_{0(2)}} = -\tensor{W}{^{(2)}_{0(1)}}$$
Applying the general field equations \eqref{Field Eqs Metric 1} and \eqref{Field 
Eqs Metric 2} for a nontrivial 
nonlinear connection in the case of the metric \eqref{bundle metric}, and 
taking 
into account the above relations, we finally obtain:
\begin{equation}\label{Friedmann 1 final}
    3H^2 + 3H \left(\frac{\dot \phi}{\phi} - W_+ \right) - W_+
    \frac{\dot\phi}{\phi} + \frac{\dot\phi^2}{4\phi^2} = 8\pi G \rho_m
\end{equation}
\begin{eqnarray}
     && \!\!\!\!\!\!\!\!\!\!\!\!\! 2\dot H + (W_+)^2 - \dot{W_+} - 
\frac{\dot\phi^2}{2\phi^2} + H 
    \left(W_+ - \frac{\dot\phi}{\phi} \right) \nonumber\\
      && \quad  \ \ \ \ \ \ \ \ -\frac{1}{2\phi}\left(W_+ \dot\phi - 2\ddot\phi 
\right) =
    - 8\pi G (\rho_m + P_m) \label{Friedmann 2 final}
\end{eqnarray}
and
\begin{eqnarray}
\label{Klein Gordon final}
      && \!\!\!\!\!\!\!\!\!\!\!\!\!\frac{1}{\phi} \left(\ddot\phi + 3H\dot\phi 
\right)- 
    \frac{\dot\phi}{2\phi}\left(3 W_+ + \frac{\dot\phi}{\phi} \right) + 6 
\left(\dot H + 2H^2\right) \nonumber\\
 && \quad  \ \ \ \ \ \ \ \ \ \ \ \  - 6H W_+ + 2 (W_+)^2 - 2 \dot{ W_+} = 
    -8\pi G \overset{V}{\mathcal T}
\end{eqnarray}
where we have defined
\begin{equation}\label{W+}
     W_+ = \tensor {W}{^a_{0a}}
\end{equation}
These are the two modified Friedmann equations and the scalar-field 
(Klein-Gordon) equation, for the scenario at hand. Indeed, as we can see we do 
obtain generalized scalar-tensor theories from  the specific  vector bundle 
model that we have constructed. 
Note, that 
according to \eqref{EM gr limit} and \eqref{perf fluid},
in the General Relativity limit we have, 
$$\overset{V}{\mathcal T} = 
-\rho_m + 3P_m$$
Therefore in the 
general case we can consider the trace as 
$$\overset{V}{\mathcal T} = -\rho_m + 3P_m + \tilde{\mathcal T}$$ where we explicitly see that  $\tilde{\mathcal T}$ is a correction
over 
the GR limit.

\subsection{  Dark Energy}

Let us now proceed to the investigation of the modified Friedmann equations 
(\ref{Friedmann 1 final}),(\ref{Friedmann 2 final}). Observing their form, we 
deduce that we can write them in the standard way, namely
\begin{align}
\label{FRW equations with effective matter1}
    3H^2 & = 8\pi G \left( \rho_m + \rho_{eff} \right) \\
    2\dot H & = -8\pi G \left( \rho_m + \rho_{eff} + P_m + P_{eff} \right) 
\label{FRW equations with effective matter2}
\end{align}
having defined an effective dark energy sector with energy density and pressure 
respectively as
\begin{align}
\label{rhoeffdef}
    \rho_{eff} & = \frac{1}{8\pi G} \Big[ \frac{\dot\phi}{\phi}W_+ - 
\frac{\dot\phi^2}{4\phi^2} -3H \Big( \frac{\dot\phi}{\phi} - W_+ \Big) 
\Big] \\
       P_{eff} & = \frac{1}{8\pi G} \Big[ (W_+)^2 -\dot W_+ - 2H W_+ - 
\frac{\dot\phi^2}{4\phi^2}   \nonumber\\
       & \quad \ \ \ \ \ \ \ \, \ \ \  +   \frac{1}{2\phi}\left(4H \dot\phi - 
3W_+ \dot\phi + 
2\ddot\phi \right) \Big]
\label{peffdef}
\end{align}
Hence, the effective dark energy sector incorporates all the extra geometrical 
information that arises from the vector bundle  construction. 

We can define the equation-of-state parameter for the effective 
dark-energy sector as
\begin{equation}
    w_{eff}\equiv\frac{P_{eff}}{\rho_{eff}}
\label{weffeq}
\end{equation}
According to the definitions (\ref{rhoeffdef}),(\ref{peffdef}), we can see that 
$w_{eff}$ can lie in the quintessence ($w_{eff}>-1$) or in the phantom  
($w_{eff}<-1$) regime, or experience the phantom-divide crossing during the 
evolution. The fact that we can effectively obtain a phantom behavior without
imposing by hand phantom fields, is an advantage of the scenario and reveals 
the capabilities of the bundle constructions. Note that  $w_{eff}$ can be even 
exactly equal to  $-1$  if one imposes the specific condition 
\beqa
\dot\psi+\frac{1}{2}\psi^2-H\psi-\frac{1}{2}W_+\psi=
\dot W_+-W_+^2-HW_+
\eeqa
where $\psi\equiv\frac{\dot{{\phi}}}{\phi}$, in which case we obtain a 
cosmological constant of effective origin, although our initial action does not 
contain an effective cosmological constant.

Finally, using the above definitions we can  examine the validity of the energy 
conditions: 
\begin{itemize}
\item{
Weak:
$\rho_{eff}\ge 0$, $\rho_{eff}+P_{eff}\ge 0$ 
}
\item{Strong:
$\rho_{eff}+P_{eff}\ge 0$, $\rho_{eff}+3P_{eff}\ge 0$
}
\item{Null: 
$\rho_{eff}+P_{eff}\ge 0$  }
\item{Dominant: 
$\rho_{eff}\ge |P_{eff}|$  }
\end{itemize} 

We proceed to the specific  investigation the cosmological behavior that is 
induced from the scenario at 
hand. In particular, we elaborate the Friedmann equations (\ref{FRW equations 
with effective matter1}), (\ref{FRW equations 
with effective matter2}) numerically, and we use   
the usual expression for the
redshift    $ 1+z=1/a$ (the present scale factor is set to
$a_0=1$) as the independent variable. This expression for the redshift is justified by two points: Firstly, we consider trajectories of the form \eqref{geodesics 1}, \eqref{parallelism condition} which effectively describe classic GR geodesics, and secondly the spacetime part of the metric \eqref{bundle metric} is identical to the classic spatially flat FRW metric of GR. Moreover, we introduce the standard 
density parameters, namely $\Omega_{m}\equiv 8\pi G\rho_{m}/(3H^2)$ and  
$\Omega_{eff}\equiv\Omega_{DE}= 8\pi G\rho_{eff}/(3H^2)$, for the matter and effective dark 
energy sector respectively. Concerning  the initial conditions
we choose them in order to 
obtain $\Omega_{eff}(z=0)\equiv\Omega_{eff0}\approx0.69$  and 
$\Omega_m(z=0)\equiv\Omega_{m0}\approx0.31$  in agreement with observations
\cite{Aghanim:2018eyx}, while for  the matter sector we impose  dust equation 
of state, namely $w_m\equiv P_m/\rho_m=0$.

\begin{figure}[ht]
\hspace{-1cm} \includegraphics[scale=0.45]{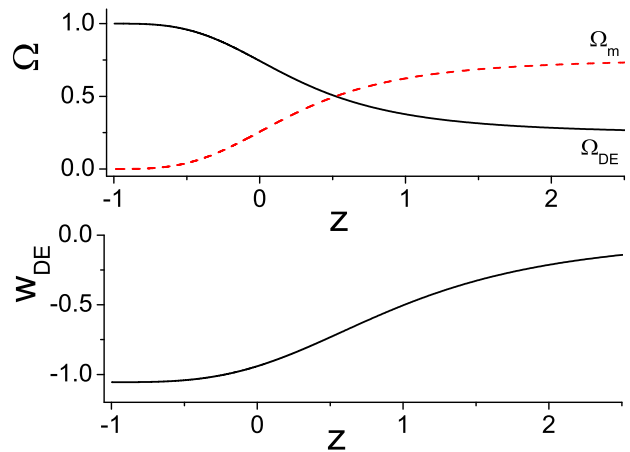}
\caption{
{\it{ Upper graph: The evolution of the effective dark energy
   density
parameter $\Omega_{DE}$ (black-solid), as well as of the
matter   
density parameter $\Omega_{m}$   (red-dashed), as a function of 
the redshift $z$. Lower graph: The evolution of the corresponding dark-energy
equation-of-state parameter $w_{DE}$.
We have imposed the initial conditions
$\Omega_{DE}(z=0)\equiv\Omega_{DE0}\approx0.69$ \cite{Aghanim:2018eyx}.
}} }
\label{OmegasFinsler}
\end{figure}

\begin{figure}[!]
\hspace{-1cm}\includegraphics[scale=0.46]{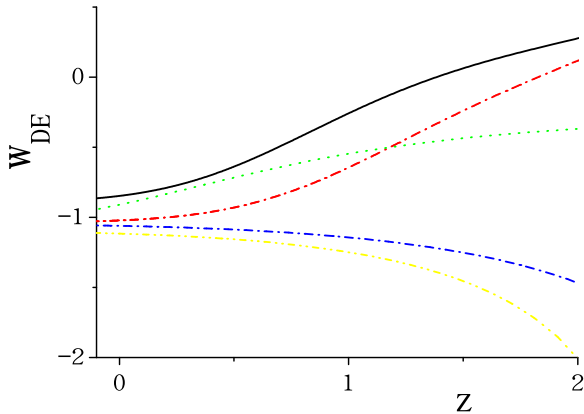}
\caption{
{\it{ The evolution of the   dark-energy
equation-of-state parameter $w_{DE}$ as a function of 
the redshift $z$, for   various small 
corrections $\tilde{\mathcal T}$. 
We have imposed the initial conditions
$\Omega_{DE}(z=0)\equiv\Omega_{DE0}\approx0.69$ \cite{Aghanim:2018eyx}.
}} }
\label{wFinsler}
\end{figure}

In the upper graph Fig. \ref{OmegasFinsler} we present 
$\Omega_{DE}(z)$
and 
$\Omega_{m}(z)$ where we observe that  we   obtain the standard 
thermal history of the universe, namely  the  matter and dark energy 
eras. Additionally, in the lower graph Fig. \ref{OmegasFinsler} we depict 
the effective dark-energy equation-of-state parameter
$w_{eff}\equiv w_{DE}$, where we can see that in this specific example the effective dark 
energy sector experiences the phantom-divide crossing during the cosmological 
evolution. In order to examine in more detail the behavior of $w_{DE}$, in 
 Fig. \ref{wFinsler} we present its evolution for various small 
corrections $\tilde{\mathcal T}$. As we can see, we can obtain a rich behavior, 
and an effective dark energy sector that can be quintessence-like,  
phantom-like, or experience the 
phantom-divide crossing. These properties cannot be easily acquired in the 
usual scalar-tensor theories, and this reveals the capabilities of the 
construction at hand.

Let us examine in more detail the conservation equations in the scenario at 
hand.  
 As expected, the 
  energy densities and pressure appearing in the Friedmann equations (\ref{FRW 
equations 
with effective matter1}), (\ref{FRW equations 
with effective matter2})  satisfy the    continuity equation 
\begin{equation}
    \dot\rho_m + \dot\rho_{eff} + 3H \left( \rho_m + \rho_{eff} + P_m + 
P_{eff}\right) = 0
\label{FRW total continuity equation}
\end{equation}
Using (\ref{Klein Gordon final}) we can re-write it as
 \begin{equation}\label{FRW continuity equation}
    \dot\rho_m + 3H(\rho_m + P_m) + \frac{\dot\phi}{2\phi}(\rho_m + 3P_m + 
\tilde{\mathcal T}) = - 
\frac{1}{8\pi G}Q_0
\end{equation}
where the time component of the dissipation vector \eqref{dissipation vector} is 
calculated as
\begin{eqnarray}
  && 
  \!\!\!\!\!\!
  Q_0 =    \frac{1}{4\phi^2}\left[12H^2\phi^2 + 12\phi^2 \dot H + 6H\phi 
    \left(2  W_+ \phi - \dot\phi \right) \right. \nonumber\\
   &  &  \ \ \ \ \ \ \ \ \ \,\ \ \ \ \ \ \left. - 3\dot\phi^2 + 4 \phi 
\left( W_+ \dot\phi + \ddot\phi \right) 
\right] 
W_+ \label{FRW dissipation vector}
\end{eqnarray}
Note that this equation can also be obtained from the time component of 
\eqref{H Delta and Q definitions} (all other components of \eqref{H Delta 
and Q definitions} give trivial equations). 
The dissipation vector  encodes 
 the energy-momentum tensor potential non-conservation 
 with respect to the 
special connection of $F^6$.  An interesting observation is that in the absence 
of matter, equations \eqref{Friedmann 1 final}, \eqref{Friedmann 2 final} and 
\eqref{Klein Gordon final} are independent, 
contrary to the standard scalar-tensor models where only two out of the three 
equations are.
Therefore, in the 
absence of 
matter, Eq. \eqref{FRW continuity equation} implies that $Q_0$ should 
vanish, a condition that makes \eqref{Klein Gordon final} dependent on 
\eqref{Friedmann 1 final} and \eqref{Friedmann 2 final}, which is then a 
 self-consistency verification of the scenario. 

In the general case the combination of 
\eqref{Friedmann 1 final} 
and 
\eqref{Friedmann 2 final} does not reproduce \eqref{Klein Gordon final}, 
exactly due to $Q_0$.
However, observing the form of \eqref{FRW continuity equation}, we deduce that 
 if we define
\begin{equation}\label{Qtilde}
    \tilde Q \equiv  -\frac{\dot\phi}{2\phi}(\rho_m + 3P_m + \tilde{\mathcal 
T}) - \frac{1}{8\pi G}Q_0
\end{equation}
then   \eqref{FRW total continuity equation} and \eqref{FRW continuity 
equation} can be re-written as
\begin{align}
    \dot\rho_m + 3H(\rho_m + P_m) & =   \tilde Q\\
    \dot\rho_{eff} + 3H(\rho_{eff} + P_{eff}) & = - \tilde Q
\end{align}
As we can see,  $\tilde Q$ represents the interaction rate between 
matter and effective dark energy sector, which lies at the basis of the matter 
non-conservation \cite{Papagiannopoulos:2017whb,Basilakos:2020}. Therefore, in the general case the scenario at hand exhibits 
an interaction between the matter component and the dark energy sector that 
quantifies the novel geometric structure of the  vector bundle. This reveals 
the capabilities of the model, since interacting cosmology is known to lead to 
very rich phenomenology \cite{Zimdahl:2001ar,Wang:2006qw,Chen:2008ft,Salvatelli:2014zta,Buen-Abad:2017gxg}
  and amongst others it can alleviate the coincidence 
problem  \cite{Zimdahl:2001ar,Jamil:2009eb} as well as the $H_0$ tension 
\cite{Pan:2019gop,DiValentino:2020zio}. However, we 
stress that in the scenario at hand the interaction between the dark sectors is 
not imposed by hand, but it naturally arises from the 
intrinsic geometrical structure of the bundle construction. 
Finally, in the particular case where  $\tilde{Q}=0$, we obtain
conservation of matter and effective dark energy sectors, i.e. we obtain the 
standard, non-interacting, cosmology.

We close this subsection by examining the special case where the condition 
$\partial_{(1)}N^{(1)}_0 = -\partial_{(2)}N^{(2)}_0 $ is imposed on the 
non-linear connection, which leads to  $W_+ = 0 $. This is 
also true when $N^a_0$ is constant, which is a solution of the Palatini field 
equations. In such a case, the modified Friedmann equations \eqref{Friedmann 1 
final}, \eqref{Friedmann 2 final}  become 
\begin{equation}\label{Friedmann 1 special}
    3H^2 + 3H \frac{\dot \phi}{\phi} + \frac{\dot\phi^2}{4\phi^2} = 8\pi G 
\rho_m
\end{equation}
\begin{equation}
    2\dot H  - \frac{\dot\phi^2}{2\phi^2} - H \frac{\dot\phi}{\phi} + 
\frac{\ddot\phi}{\phi} = - 8\pi G (\rho_m + P_m) \label{Friedmann 2 special}
\end{equation}
while the  Klein-Gordon equation \eqref{Klein Gordon final}  is simplified to
\begin{equation}
    \frac{1}{\phi} \left(\ddot\phi + 3H\dot\phi \right) - 
\frac{\dot\phi^2}{2\phi^2} + 6 \left(\dot H + 2H^2\right) =
8\pi G (\rho_m - 3P_m - \tilde{\mathcal T}) \label{Klein Gordon special}
\end{equation}
Note that the interaction between matter and effective dark energy sector is 
maintained.

\subsection{Cold Dark Matter}

One of the features of the construction at hand is that  the metric 
 of the base manifold can be decomposed into an ordinary  and a 
 dark  matter piece  according to (\ref{base metric separation}). As a result, 
the perfect fluid \eqref{perf fluid} can be decomposed into ordinary and cold 
dark matter (CDM) sectors \cite{Savvopoulos:2020}, which using   
Eq. \eqref{Tmn split}
leads to
\begin{align}
    \mathcal T_{\mu\nu} & = \left(\rho_m^{(O)}+ P_m^{(O)} \right) Y_\mu Y_\nu + 
P_m^{(O)} \left(\tensor{g}{^{(O)}_{\mu\nu}}(x) + \tensor{g}{^{(D)}_{\mu\nu}}(x) 
\right) \nonumber\\
    & \, + \rho_m^{(D)} Y_\mu Y_\nu \label{FRW energy-momentum split}
\end{align}
Note that we have assumed that, due to spatial isotropy, the ordinary 
matter and  CDM fluids are at rest with respect to the comoving grid, and thus 
they have the same 4-velocity $Y^\mu$, and moreover that the CDM fluid is 
pressureless as usual ($P_m^{(D)} = 0$).
Expression \eqref{FRW energy-momentum split} can be decomposed into ordinary, 
dark and interaction terms, respectively as
\begin{align}
    \tensor{\mathcal T}{^{(O)}_{\mu\nu}} & = \left(\rho_m^{(O)}+ P_m^{(O)} 
\right) Y_\mu Y_\nu + P_m^{(O)} \tensor{g}{^{(O)}_{\mu\nu}}(x) \\
    \tensor{\mathcal T}{^{(D)}_{\mu\nu}} & = \rho_m^{(D)} Y_\mu Y_\nu \\
    \tensor{\tau}{_{\mu\nu}} & = P_m^{(O)} \tensor{g}{^{(D)}_{\mu\nu}}(x)
\end{align}
In this case,  the modified  Friedmann 
equations  \eqref{FRW equations with effective matter1}, \eqref{FRW equations 
with effective matter2}     take the form
\begin{align}
    3H^2 & = 8\pi G \left[ \rho^{(O)}_m + \rho^{(D)}_m + \rho_{eff}\right]
    \label{FRW equation with dark sector 1} \\
    2\dot H & = -8\pi G \left[ \rho^{(O)}_m + \rho^{(D)}_m+ \rho_{eff} + 
P^{(O)}_m + P_{eff} \right]
\label{FRW equation with dark sector 2}
\end{align}
Additionally, the continuity equation \eqref{FRW continuity equation}  becomes
\begin{align}
    & \dot\rho^{(O)}_m + \dot\rho^{(D)}_m  + 3H \left(\rho^{(O)}_m + 
\rho^{(D)}_m  + P^{(O)}_m \right) \nonumber\\
    & \ \ \ \quad = - \frac{\dot\phi}{2\phi} \left(\rho^{(O)}_m + \rho^{(D)}_m 
+ 3P^{(O)}_m + \tilde{\mathcal T}\right) - \frac{1}{8\pi G}Q_0 \label{FRW 
continuity equation CDM}
\end{align}
We observe that this relation provides an effective source term  with respect 
to General Relativity. This term can be traced to the fibre 
components of our special connection, which provide the first term on the 
right hand side, and to the dissipation term $Q_0$, hence to the 
non-conservation of the energy-momentum tensor with respect to the connection 
of 
$F^6$. Focusing on the CDM sector, assuming that the dark matter content is 
close to its GR limit ($\tilde{\mathcal T} \approx 0$), and considering the 
special case where   $W_+$ vanishes, which according to   \eqref{FRW 
dissipation vector} leads to $Q_0 = 0$,   equation \eqref{FRW 
continuity equation CDM} becomes: 
\begin{equation}\label{CDM creation in F6}
    \dot \rho^{(D)}_m + 3H \rho^{(D)}_m = - \frac{\dot\phi}{2\phi} \rho^{(D)}_m
\end{equation}
Observing equation \eqref{CDM creation in F6}, we find a parallelism with 
models 
of CDM creation in GR \cite{Capozziello:2008it}. In particular, the continuity 
equation of CDM in these 
models reads \cite{Lima:2007kk,Lima:2008qy,Lima:2012cm,Basilakos:2020}:
\begin{equation}\label{CDM creation equation}
    \dot \rho^{(D)}_m + 3H \rho^{(D)}_m = \Gamma \rho^{(D)}_m
\end{equation}
where $\Gamma$ is the CDM creation rate. Comparing \eqref{CDM creation in F6} 
with \eqref{CDM creation equation}, we find that our model provides a dynamics 
for CDM creation similar to the aforementioned models, namely
\begin{equation}
\label{creation rate}
    \Gamma = - \dfrac{\dot\phi}{2\phi}
\end{equation}
From the point of view of an observer who interprets the creation mechanism 
in the framework of General Relativity and standard FRW cosmology, it would 
appear that \eqref{CDM creation in F6} violates the conservation of 
energy-momentum due the appearance of the source term in the rhs. However, from 
the point of view of our construction, the same mechanism can be seen as a 
result of energy-momentum conservation with respect to the special connection 
of 
the total space $TF^6$. Once again we mention that this behavior has not be 
imposed  by hand, but it  arises naturally from the 
  geometrical structure of the bundle construction.

\section{Concluding Remarks}
\label{Conclusions}

In this article we studied 
the gravitational and cosmological consequences of a, Finsler-like, scalar 
tensor theory on a vector bundle $F^6$, which consists of a pseudo-Riemannian 
space-time manifold with two scalars in the role of fibres or internal 
variables. In this approach, we used a non-linear connection form of a 
non-holonomic bundle structure. Under this framework,
the properties of a 
sectorized gravitational field are analyzed for both the ordinary and dark 
sectors.

The extra geometrical structure is imprinted in the field equations 
\eqref{Metric Field Eqs Bundle}, Raychaudhuri \eqref{Raychadhuri generalisation} 
and FRW equations \eqref{Friedmann 1 final}, \eqref{Friedmann 2 final}, 
\eqref{Klein Gordon final}. Due to the introduction of the scalar fields 
$\phi^{(1)}$, $\phi^{(2)}$ we obtain extra  degrees of freedom
which affect the volume of congruence geodesics, the form of the accelerating 
universe and potentially lead to 
Lorentz violating and locally anisotropic effects 
\cite{Kostelecky,Kostelecky-Russel, Bastero, Kanno, Almeida}. 
An interesting topic for the upcoming projects would be to examine whether the symmetries of usual spatial homogeneity and  
isotropy persist on the vector bundle $E$. A careful and meaningful definition of these symmetries should be given in the current framework and the most consistent way to achieve this is through the proper extension of the concepts of Lie derivatives and Killing vector fields. We remark that the kind of isotropy we are discussing here differs from the concept of internal
spacetime anisotropy encountered in Finsler gravity.

Applying this construction at a cosmological framework, we showed that the 
induced generalized scalar-tensor theory from the bundle structure and the 
non-linear connection leads to the appearance of an effective dark energy 
sector in the modified Friedmann equations. Hence, we were able to reproduce 
the thermal history of the universe, with the sequence of the matter and dark 
energy eras, and  we showed that the resulting dark-energy 
equation-of-state parameter can lie in the quintessence or phantom regime, or 
even exhibit the phantom-divide crossing. Furthermore, we showed that this 
novel intrinsic geometrical structure leads to an effective interaction between 
the dark matter and the metric
and for the particular
case of cold dark matter
the  relation \eqref{creation rate}  was found  between the scalar fields and the CDM creation rate.

There are many things that one should do in order to further investigate 
generalized scalar-tensor theories arising from vector bundle constructions. 
The first is to study the specifically symmetric and black hole solutions, and 
examine the differences comparing to General Relativity. The second is to 
consider specific examples of non-linear connections and examine whether they 
can lead to distinguishable behavior. Finally, one should 
investigate  in more detail the cosmological applications, incorporating  
data from  Type Ia Supernovae (SNIa), Baryon Acoustic Oscillations (BAO), 
Cosmic Microwave 
Background (CMB) observations.   These interesting and necessary studies   are 
left for future projects.

\section*{Acknowledgments}
The authors would like to thank the uknown referee for his/her valuable comments and remarks. 
This research is co-financed by Greece and the European Union (European Social 
Fund-ESF) 
through the Operational Programme ``Human Resources Development, Education and 
Lifelong 
Learning'' in the context of the project ``Strengthening Human Resources 
Research 
Potential via Doctorate Research'' (MIS-5000432), implemented by the State 
Scholarships 
Foundation (IKY). This article is based upon work from COST Action CA18108 
``Quantum Gravity Phenomenology in the multi-messenger approach'', supported by 
COST (European Cooperation in
Science and Technology).

\begin{appendix}

\section{Connection and curvature}\label{Connection and curvature}
One can define a special type of linear connection in this space, where the 
following 
rules 
hold:
\begin{align}
	& D_{\delta_\nu}\delta_\mu = L^\kappa_{\mu\nu}\delta_\kappa    & 
	&D_{\delta_\nu} \partial_{a} = L^{c}_{{a}\nu} 
	 \partial_{c} \label{Dh} \lin
	& D_{ \partial_{b}} \delta_\mu = 
	C^{c}_{\mu b} \partial_{c}   & 
	&D_{ \partial_{b}} \partial_{a} = 
	C^\kappa_{ab}\delta_\kappa \label{Dv}
\end{align}
Differentiation of the inner product $D_{X_K}<X^M,X_N> = 0$ and use of 
\eqref{Dh},\eqref{Dv} leads to the rules:
\begin{align}
	& D_{\delta_\nu}\de x^\kappa = -L^\kappa_{\mu\nu}\de x^\mu &  
	&D_{\delta_\nu}\delta\phi^{c} = - L^{c}_{{a}\nu} 
	\delta\phi^{a} \lin
	& D_{ \partial_{b}} \de x^\kappa = - C^\kappa_{ab}\delta 
	\phi^{a}\! &  &D_{ \partial_{b}}\delta\phi^{c} = - 
	C^{c}_{\mu b} \de x^\mu
\end{align}
It is apparent from the above relations that $D_{\delta_\nu}$ preserves the 
horizontal 
and 
vertical distributions, while $D_{ \partial_{b}}$ maps one to the other.

Following the above rules, covariant differentiation of a vector $V = 
V^\mu\delta_\mu + 
V^{a} \partial_{a}$ along a horizontal direction gives:
\begin{align}
	D_{\delta_\nu}V & = \left(\delta_\nu V^\mu + V^\kappa L^\mu_{\kappa\nu} 
	\right)\delta_\mu + \left(\delta_\nu V^{a} + 
	V^{c}L^{a}_{c\nu}\right) \partial_{a} \nonumber\\
	& = D_\nu V^\mu \delta_\mu + D_\nu V^{a} \partial_{a}
\end{align}
where we have defined
\begin{align}
	& D_\nu V^\mu = \delta_\nu V^\mu + V^\kappa L^\mu_{\kappa\nu} \\
	& D_\nu V^{a} = \delta_\nu V^{a} + 
	V^{c}L^{a}_{c\nu}
\end{align}
Similarly, for the covariant differentiation of $V$ along a vertical direction 
we obtain
\begin{align}
	D_{ \partial_{b}}V  = & \left[ \partial_{b} V^\mu +  
	V^{a}C^\mu_{ab} 
	\right]\delta_\mu \nonumber\\
	&  + \left[ \partial_{b} V^{a} + V^\mu 
	C^{a}_{\mu b}\right] \partial_{a} \nonumber\\
	= &\, D_{b} V^\mu \delta_\mu + D_{b} 
	V^{a} \partial_{a}
\end{align}
where we have defined
\begin{align}
	& D_{b} V^\mu =  \partial_{b} V^\mu +  
	V^{a}C^\mu_{ab} \\
	& D_{b} V^{a} =  \partial_{b} V^{a} + V^\mu 
	C^{a}_{\mu b}
\end{align}
The covariant derivative over the full range of indices in $F^6$ reads:
\begin{align}
    D_{X_M} V & = \left[ X_M V^N + \tensor{\bs\Gamma}{^N_{LM}} V^L \right]X_N 
\nonumber\\
              & = \left(D_M V^N \right) X_N
\end{align}
where
\begin{equation}
    D_M V^N = X_M V^N + \tensor{\bs\Gamma}{^N_{LM}} V^L
\end{equation}
Finally, the covariant derivative for a tensor of general rank is obtained in a 
similar 
way.

\section{Field equations}\label{sec: Field Eqs}

In this Appendix   we present the steps which lead to the field 
equations, \eqref{Field Eqs Palatini 1}, \eqref{Field Eqs Palatini 2}, 
\eqref{Field Eqs Metric 1} and \eqref{Field Eqs Metric 2}.
A Hilbert-like action with a matter sector on the bundle $F^6$ is
\begin{equation}
    S = \int_Q d^6U \sqrt{|\mathcal G|} \mathcal R + 2\kappa \int_Q d^6U 
\sqrt{|\mathcal G|} \mathcal L_M(\mathcal G^{MN},\Psi^i)
\end{equation}
where $\mathcal L_M(\mathcal G^{MN},\Psi^i)$ is the Lagrangian of the matter fields 
$\Psi^i$, and $Q$ is a closed subspace of $F^6$. Variation of the action gives
\begin{align}
    \delta S = &\, \int_Q d^6U \sqrt{|\mathcal G|} \left( \mathcal R_{MN} - 
\frac{1}{2}\mathcal G_{MN} \mathcal R \right) \delta \mathcal G^{MN} \nonumber\\
               & + \int_Q d^6U \sqrt{|\mathcal G|} \mathcal G^{MN} \delta 
\mathcal R_{MN} 
               \nonumber\\
    &\,        + 2\kappa \int_Q d^6U \delta\left(\sqrt{|\mathcal G|} \mathcal 
L_M \right)  =  0 \label{Maction1}
\end{align}
After a straightforward calculation, we acquire
\begin{align}
	\mathcal G^{MN} \delta \mathcal R_{MN}& =  D_K \left[ \mathcal 
G^{MN}\delta\bs \Gamma^K_{MN} - \mathcal G^{\mu\nu}\bs\Gamma^K_{\mu b}  \delta 
N^b_\nu \right] \nonumber\\
	& - D_N \left[ \mathcal G^{MN}\delta\bs\Gamma^K_{MK} - \mathcal 
G^{MN}\bs\Gamma^\kappa_{Mb} \delta N^b_\kappa \right] \nonumber\\
	                      & + \mathcal G^{MN} \tensor{\mathscr 
T}{^Z_{NK}}\delta\bs\Gamma^K_{MZ} \nonumber\\
	                      & + \mathcal G^{\mu\nu}  \left[ \delta N^b_\nu 
\mathcal R_{\mu b} + \delta N^b_\kappa \left( \tensor{\mathscr 
T}{^\kappa_{A\nu}} \bs\Gamma^A_{\mu b} - \mathcal R^\kappa_{\mu\nu b}\right) 
\right]
\end{align}
Applying Stoke's theorem to the above result and \eqref{Maction1}, and assuming 
that the boundary terms vanish, leads to the following relation:
\begin{align}
	\mathcal G^{MN} \delta \mathcal R_{MN} = &\, \tensor{\mathscr T}{^A_{KA}} 
\left[ \mathcal G^{MN}\delta\bs \Gamma^K_{MN} - \mathcal 
G^{\mu\nu}\bs\Gamma^K_{\mu b}  \delta N^b_\nu \right] \nonumber\\
	& - \tensor{\mathscr T}{^A_{NA}} \left[ \mathcal 
G^{MN}\delta\bs\Gamma^K_{MK} - \mathcal G^{MN}\bs\Gamma^\kappa_{Mb} \delta 
N^b_\kappa \right] \nonumber\\
	                      & + \mathcal G^{MN} \tensor{\mathscr 
T}{^Z_{NK}}\delta\bs\Gamma^K_{MZ} \nonumber\\
	                      & + \mathcal G^{\mu\nu}  \left[ \delta N^b_\nu 
\mathcal R_{\mu b} + \delta N^b_\kappa \left( \tensor{\mathscr 
T}{^\kappa_{A\nu}} \bs\Gamma^A_{\mu b} - \mathcal R^\kappa_{\mu\nu b}\right) 
\right] \label{GDeltaR}
\end{align}
In the Palatini method, the fields $\mathcal G^{MN}$, $\bs\Gamma^L_{MN}$ and 
$N^a_{\mu}$ are varied independently from each other, therefore  
\eqref{Maction1} and \eqref{GDeltaR} provide the equations 
\begin{align}
    \mathcal R_{(MN)} - \frac{1}{2}\mathcal G_{MN} \mathcal R = \kappa \mathcal 
T_{MN}
\end{align}
\begin{equation}
    \mathcal G^{MN} \tensor{\mathscr T}{^A_{KA}} + \mathcal G^{ML}\left( 
\tensor{\mathscr T}{^N_{LK}} - \tensor{\mathscr T}{^A_{LA}}\delta^N_K \right) = 
0
\end{equation}
and
\begin{align}
    & \tensor{\mathscr T}{^A_{NA}} \mathcal G^{MN}\bs\Gamma^\kappa_{Mb} - 
\tensor{\mathscr T}{^A_{KA}}\mathcal G^{\mu\kappa}\bs\Gamma^K_{\mu b} 
\nonumber\\
    & \qquad + \mathcal G^{\mu\nu} \left( \delta^\kappa_\nu \mathcal R_{\mu b} + 
\tensor{\mathscr T}{^\kappa_{A\nu}}  \bs\Gamma^A_{\mu b} - \mathcal 
R^\kappa_{\mu\nu b}\right) = 0 \label{Field Eqs Palatini 3}
\end{align}
 where
\begin{equation}\label{energy momentum tensor}
    \mathcal T_{MN} = -\frac{2}{\sqrt{|\mathcal G|}}\frac{\delta 
\left(\sqrt{|\mathcal G|}\mathcal L_M\right)}{\delta \mathcal G^{MN}}
\end{equation}
We remark that the lhs of \eqref{Field Eqs Palatini 3} vanishes identically for 
the choice of connection \eqref{Symmetric Connection}.

Alternatively, we can variate the action by considering all the fields dependent 
on the metric $\mathcal G^{MN}$. For the specific connection components given in 
\eqref{Symmetric Connection}, the non-vanishing part of Eq.\eqref{GDeltaR} 
reads:
\begin{align}
   &\mathcal G^{MN} \delta \mathcal R_{MN} \nonumber\\
   & = \left( \delta^M_A \delta^K_B - \mathcal G^{MK} \mathcal 
G_{AB}\right)\left(D_M\tensor{\mathscr T}{^Z_{KZ}} - \tensor{\mathscr 
T}{^L_{ML}} \tensor{\mathscr T}{^Z_{KZ}}\right) \delta \mathcal G^{AB} 
\label{GDeltaR Metric}
\end{align}
where we have used Stoke's theorem twice and eliminated all the boundary terms. 
Combining \eqref{Maction1} and \eqref{GDeltaR Metric} gives Eqs. \eqref{Field 
Eqs Metric 1} and \eqref{Field Eqs Metric 2}.
\vspace{5pt}

\section{Generalised Bianchi identities}\label{sec: Bianchi}
The Bianchi identities constrain the curvature and torsion tensors via the 
relations \cite{miron-watanabe-ikeda:1987}:
\beqa
&&\label{Bianchi General 1}
\circlearrowleft_{A,M,N} \left\{D_A \tensor{\mathcal R}{^K_{LMN}} + 
\tensor{\mathcal R}{^K_{LAR}} \tensor{\mathcal W}{^R_{MN}}\right\} = 0 \\
&&\label{Bianchi General 2}  
\circlearrowleft_{A,M,N} \left\{D_A \tensor{\mathcal W}{^L_{MN}} + 
\tensor{\mathcal W}{^K_{AM}} \tensor{\mathcal W}{^L_{NK}} + \tensor{\mathcal 
R}{^L_{AMN}}\right\} = 0\nn \\
\eeqa
In our calculations we will use the symmetry
\begin{equation}
\tensor{\mathcal R}{^K_{LMN}}=- \tensor{\mathcal R}{^K_{LNM}} \label{Riemann 
symmetry 1} 
\end{equation}
which is obvious from the defining relation of the Riemann tensor 
\eqref{Riemann}. Manipulating 
\beqa
\mathcal R_{KLMN}=\mathcal G_{KR}\tensor{\mathcal R}{^R_{LMN}}
\eeqa
with the aid of \eqref{Symmetric Connection} and \eqref{Riemann}, it can be 
shown that
\beqa\label{Riemann covariant}
\mathcal R_{KLMN}&=&\frac{1}{2}\Big(X_M X_L\mathcal G_{NK}-X_M X_K\mathcal 
G_{LN}-X_NX_L\mathcal G_{MK}\nn \\
&&+X_NX_K\mathcal G_{LM}
\Big)+\tensor{\pmb{\Gamma}}{^R_{LM}}\tensor{\pmb{\Gamma}}{_{RNK}}-\tensor{\pmb{
\Gamma}}{^R_{LN}}\tensor{\pmb{\Gamma}}{_{RMK}}\nn \\&&+\tensor{\mathcal 
W}{^R_{MN}}
\tensor{\pmb{\Gamma}}{_{LRK}}
\eeqa
From the above, it can be seen that generally 
\begin{equation}\label{symmetric over first two indices}
\mathcal R_{(KL)MN}=\frac{1}{2}\tensor{\mathcal W}{^R_{MN}}X_R\mathcal G_{KL}
\end{equation}
However, it is obvious from \eqref{structure functions components} that only the 
latin upper index elements of $\mathcal W$ are non-zero. Since this is index is 
contracted with the derivative of the metric with respect to the fibre 
variables, $\mathcal R_{(KL)MN}$ is always zero. Therefore, we deduce the 
antisymmetry of $\mathcal R_{KLMN}$ with respect to its
first two indices. i.e. 
\beqa\label{Riemann symmetry 2}
\mathcal R_{KLMN} = -\mathcal R_{LKMN}
\eeqa
Again from \eqref{Riemann covariant},
and taking advantage of \eqref{symmetric over first two indices}
and \eqref{Riemann symmetry 2}, we prove
\beqa\label{2nd Bianchi simplified}
\circlearrowleft_{L,M,N}\{\mathcal R_{KLMN}+\tensor{\mathcal 
W}{^R_{MN}}\tensor{\pmb{\Gamma}}{_{KRL}}\}=0
\eeqa
which is equivalent to \eqref{Bianchi General 2}, and 
\beqa
\mathcal R_{KLMN}=\mathcal R_{MNKL}+
\tensor{\mathcal W}{^R_{MN}}\tensor{\pmb{\Gamma}}{_{LRK}}
-\tensor{\mathcal W}{^R_{KL}}\tensor{\pmb{\Gamma}}{_{NRM}}\nn  
\eeqa
Lastly, from the above equation we can derive the symmetry properties of the 
generalised Ricci tensor:
\beqa\label{Ricci antisymmetric}
\mathcal R_{MN}=\mathcal R_{NM}+
\tensor{\mathcal W}{^L_{RM}}\tensor{\pmb{\Gamma}}{^R_{LN}}-
\tensor{\mathcal W}{^L_{RN}}\tensor{\pmb{\Gamma}}{^R_{LM}}
\eeqa
From the first identity \eqref{Bianchi General 1} we have:
\begin{align}
	 &\,
	 \tensor{\mathcal G}{^{AL}}
	 \tensor{\mathcal G}{^M_K} \left(D_A
	\tensor{
	\mathcal R}{^K_{LMN}} + D_M\tensor{
	\mathcal R}{^K_{LNA}} +D_N\tensor{
	\mathcal R}{^K_{LAM}} \right. \nonumber\\
	&  \left. + 
	\tensor{
	\mathcal R}{^K_{LAR}}
	 \tensor{\mathcal W}{^R_{MN}} + \tensor{
	\mathcal R}{^K_{LMR}}
	 \tensor{\mathcal W}{^R_{NA}} + \tensor{
	\mathcal R}{^K_{LNR}}
	 \tensor{\mathcal W}{^R_{AM}} \right) = 0\nn
	 \end{align}
and after some algebra we finally obtain
\beqa	 
 D^A\mathcal E_{AN}+\tensor{\mathcal R}{^A_R}
\tensor{\mathcal W}{^R_{NA}}+
\frac{1}{2}\tensor{\mathcal R}{^{KA}_{NR}}
\tensor{\mathcal W}{^R_{AK}}=0 \label{Bianchi}
\eeqa
\end{appendix}

\end{document}